\documentclass[12pt]{revtex4}
\usepackage[english]{babel}
\usepackage[utf8x]{inputenc}
\usepackage{amsmath}
\usepackage{graphicx}
\usepackage[colorlinks=true, allcolors=blue]{hyperref}
\usepackage{caption}
\usepackage{subcaption}

\begin{document}

\title{Modeling an urban highway: a statistical physics point of view for a nonphysical system}
\author{Leonardo Castro Gonz\'alez}
\email{leonardo\_castro@ciencias.unam.mx}
\affiliation{Centro de Ciencias de la Complejidad, Universidad Nacional Autónoma de México\\ Circuito Centro Cultural S/N, Ciudad Universitaria, Delegaci\'on Coyoac\'an, 04510, Ciudad de M\'exico, M\'exico }
\author{Mar\'ia Elena L\'arraga}
\email{mlarragar@iingen.unam.mx}
\affiliation{Instituto de Ingernier\'ia, Universidad Nacional Aut\'onoma de M\'exico\\ Cirucito exterior S/N, Ciudad Universitaria \\Delegaci\'on Coyoac\'an\\ 04510, Ciudad de M\'exico, M\'exico}
\author{J. Antonio del R\'io}
\email{arp@ier.unam.mx}
\affiliation{Instituto de Energía Renovables. Universidad Nacional Aut\'onoma de M\'exico\\  Privada Xochicalco S/N, 62580, Temixco, Morelos, M\'exico}
\affiliation{Centro de Ciencias de la Complejidad, Universidad Nacional Autónoma de México\\ Circuito Centro Cultural S/N, Ciudad Universitaria, Delegaci\'on Coyoac\'an, 04510, Ciudad de M\'exico, M\'exico }

\begin{abstract}
Nowadays, methodologies coming from studying physical systems are being applied to the description of a wide variety of complex systems. In particular, one can study thermodynamical methods to describe the overall behavior of many systems, independent of the precise microscopic construction. In this paper, a real Mexican highway is studied as a cellular automata system using available official data released by the Mexican Government. The system studied is the Cuernavaca bypass which was modified in 2016. Official data allows to compare the highway before and after the modifications. As more complex thermodynamic variables such as entropy is difficult to define and measure in discrete traffic models, it is shown how other more simple variables such as the standard deviation can be enough to have a complete analysis of the system. More specifically, it is shown how standard deviation can be seen as a measure of order. Results from the study of the highway show how, taking a minimal measure such as ordering the transit of heavy trucks can reduce up to 32\% the travel time from one end to another. Otherwise, travel times stays practically constant with respect to the original system. 
\end{abstract}
\pacs{45.70.Vn, 89.40.-a, 05.10.-a}

\maketitle

\section{Introduction}

Nowadays physics are contributing to methodologies, which have been successfully used for decades to address problems of many bodies, describing phenomena of social, economic and biological systems. In general, the methodologies of physics are being applied to the description of a wide variety of complex systems. Among those, a cellular automaton-based description has become most fruitful due to its the relative simplicity and flexibility.

For instance, a few decades ago Sznajd-Weron \cite{sznajd, Galam} proposed an Ising model to model the decision-making processes in parliaments. The idea is straightforward, members of the parliament raise their hands driven by the interactions with their neighbors and an external pressure, due to public opinion or to a decision of the party to which they belong. For Sznajd, the parliamentary dynamics seemed similar to the behavior of magnetic particles interacting with each other and with the presence of an external field.

On the other hand, in a pioneering work, Montroll \textit{et al.} \cite{Reiss1986} show how thermodynamics could describe non-physical systems and addressed, in particular, the example of vehicular traffic. The cited work illustrates how the standard deviation of an average quantity can be associated with the entropy of the system. Nowadays, one can study thermodynamics to describe the overall behavior of many systems, independent of the precise microscopic construction of each system using simple models, which are more amenable to efficient simulation, and potentially to statistical analysis. The fact that the dynamics of cellular automata (CA) can be implemented in the form of intuitive rules has allowed to include rather complex aspects of the behavior of the part in a rather simple way \cite{Andreas2009}.

Cellular automata are recognized as a simple modeling paradigm, which offers to graduate students an alternative method to the commonly used analytical approach, in order to study many complex systems. In this type of models attention is paid explicitly to each individual integrating the studied system and to the interactions among these generated by the the mutual influence \cite{Andreas2009}. Space is discrete and consists of a regular grid of cells, where each cell is in a particular state belonging a finite set of states. All cell states are updated synchronously in discrete time steps. Updating obeys a finite set of local interaction rules that can have a probabilistic influence. The new state of a cell is determined by the actual state of the cell itself and its neighbor cells. This local interaction allows to capture micro-level dynamics and propagates it to macro-level behavior. Thus, CA can be viewed as discrete approximations to particles dynamics and transport phenomena.

The aim of this work is multiple. Our general objective is to study a real highway as a cellular automata system using available official data released by the Mexican Government while using statistical physics tools. The highway studied is the Cuernavaca bypass: 27.3 kilometers crossing the southern Mexican city of Cuernavaca. The highway is not only important in a local metropolitan context, but also in the transport of goods between the Pacific Ocean and Mexico City. The bypass was modified, adding lanes to the existing ones and modifying the topology of the highway. However, questions have arisen in present times about its functionality, mainly because of construction problems finishing in a meters-wide crack in the middle of the highway.
In that sense, the main question to be answered in the present work is: do the modifications to the Cuernavaca bypass improve the mobility of vehicles compared to the original highway?

The model used is based in the Nagel-Schreckenberg model\cite{NaSch}, where the cells of the highway can be occupied or not by a vehicle. As space and time is discrete, so it is speed. A maximum speed is imposed.
To study the system, basic tools such as the standard deviation of the speed are used to describe the behavior of vehicles on a highway. Indeed this macroscopic measure can be used to detect phase transitions in the system when other macroscopic variables such as entropy or temperature do not appear as ``organically''. 

In Section \ref{sec:Highway}, the studied highway and its modifications are presented. Then, in Section \ref{sec:Model} the model used is presented. Available data processed and used to analyze the system is presented in Section \ref{ssec:Data}. A short introduction to the different phases that are found in traffic systems and how they can be linked with physics is done in Section \ref{sec:Methods}. Also, in the latter section the methods to use the standard deviation as a measure of order are presented. Finally, results and conclusions are shown in Section \ref{sec:Results} and \ref{sec:Conclusions} respectively.

\section{Characteristics of the Highway\label{sec:Highway}}

\begin{figure}[ht]
\centering
\includegraphics[width=0.4\textwidth]{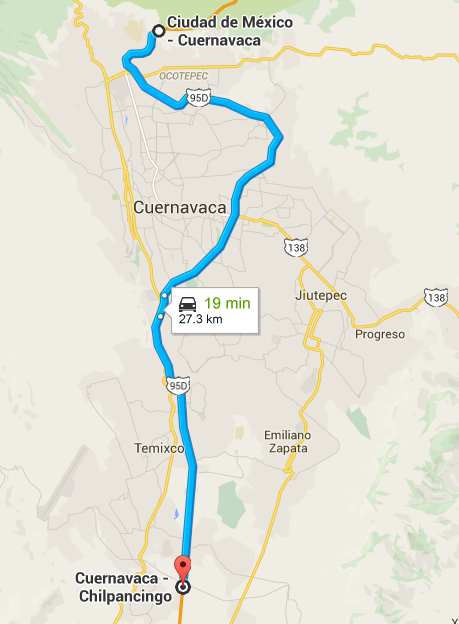}
\caption{Map of the Cuernavaca's bypass. The modification goes from the northern end to the point where the white window starts. Obtained from Google Maps; Map Data: \copyright 2018 Google, INEGI.}
\label{fig:map}
\end{figure}

The idea of the Cuernavaca bypass is to have a highway that, as said, allows to cross in a North-South direction the city, thus having a double objective: to allow local traffic of Cuernavaca to have a quick way to transport themselves within the city, and to allow the transportation of vehicles going from Mexico City to the Pacific shore. In that sense, the bypass is part of a larger highway with ends at Mexico City and Acapulco. Governmental modifications only consider the 14.5 kilometers of the bypass crossing Cuernavaca. However, as the metropolitan zone receives local traffic from other counties surrounding Cuernavaca, the studied system is extended by 12.8 kilometers to the south, resulting in a total of 27.3 kilometers. Governmental modifications are exposed in Section \ref{ssec:modifications}.

The North-South sense of the highway is called S1. Before the modifications, the road was composed of two lanes. The South-North sense is designated as S2. Originally, S2 was also a two lanes road for the first 26 km. In the last 1.3 km, an additional lane was added to the right of the highway.

\subsection{\label{ssec:slope}Topographical elements}
Along the highway, two different topographical elements reduce the maximum speed of vehicles. First, dangerous curves at the geographical north of the system diminish maximum speed to approx. $v_\mathrm{max}=100 $km/h in both senses S1 and S2. 
These curves are located along the first (last) two and a half kilometers of S1 (S2) and two kilometers after (before). The curves can be located in Figure \ref{fig:map}. In the northern part of the highway, a humanoid face is delimited by the road. Dangerous curves are here the chin and the nose of that face.

The mountains surrounding Cuernavaca at the north create a continuous slope which affects heavy transportation when driving north. Slope has such an importance that their maximum speed is diminished to approx. $v_\mathrm{max}=60$ km/h. The effect on the slow vehicles by the slope is only modeled at the northern third of the bypass and only applies to S2.

\subsection{\label{ssec:modifications}Modifications done to the highway}
\begin{figure}[ht]
\centering
\includegraphics[width=0.7\textwidth]{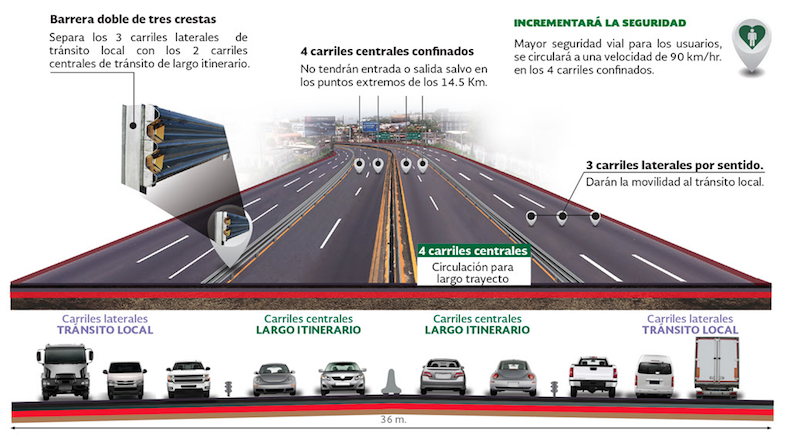}
\caption{\label{fig:ampliacion}Official scheme of the modifications made to the Cuernavaca bypass (in Spanish).}
\end{figure}

The modifications done by the Mexican Government can be summed up in two: the addition of three lanes to each sense, resulting in a highway of ten lanes taking into account both senses, and the creation of an ``express pass'' in the four middle lanes (two lanes in each sense) where no ramp allowing any car to enter or exit exists. A scheme of this is presented in Figure \ref{fig:ampliacion}. This results in the distinction of three different systems for analyzing in each direction:
\begin{itemize}
\item S1 Original - two lanes during 27.3km with 11 ramps;
\item S1 Express Pass - two lanes during 27.3km with 5 ramps;
\item S1 Local Traffic - three lanes during 14.5km and two lanes during the last 12.8km with 12 ramps;
\item S2 Original - two lanes during 16km and three lanes during the last 1.3km with 10 ramps;
\item S2 Express Pass - two lanes during 27.3km with 5 ramps;
\item S2 Local Traffic - two lanes during first 12.8km and three lanes during last 14.5km with 10 ramps. 
\end{itemize}

\section{\label{sec:Model}Model}
An already studied Cellular Automata model \cite{NaSch,Larraga20105425,delRio, Schad1} has been  used to analyze the Mexico City-Cuernavaca highway.  However, the past studies focused on the non urban part of the highway where only one ramp is present.

Even if, without a doubt, there are more complex and accurate traffic models \cite{ca6,ca10,Book1,GLA-JCA2015,GUZMAN2018}, the model used here was chosen because of two reasons: the simplicity of the model makes it a minimal one. In that sense, there are only a small number of parameter to adjust and the rules it follows makes it very intuitive and straightforward to use. Also, the results it reproduces are in accord with the highways observed in daily life \cite{Schad1,larragadelrio2005}.

\subsection*{\label{ssec:NaSch}NaSch model with anticipation parameter}

A Cellular Automata model based in the Nagel-Schreckenber (NaSch) model \cite{NaSch} and modified to have an anticipation parameter \cite{Schad1} is used. The modifications are fully detailed in \cite{delRio, Schad1,LarragaDoc}. A discrete array of $L$ cells and open boundaries is used. Each cell is considered to have a length of $7.5$m, and is rather used by a vehicle or not. Each vehicle has a discrete position $x \in \{0, 1,\ldots, L\}$ and speed $v \in \{0, 1, \dotsc, v_{\mathrm{max}}\}$. The time step is considered to be $1$ second.

Two types of vehicles are used: a fast one representing common 3/5-doors cars in good conditions and a slow one representing heavy vehicles such as trucks or buses. Fast vehicles are denoted as type 2 vehicles. They have a maximum speed of $v_{\mathrm{max}} = 5 \ \mathrm{cells}/\mathrm{s}=135 \ \mathrm{km/h}$ and a length of one cell. Slow vehicles are denoted as type 1, having a maximum speed of $v_{\mathrm{max}} = 3 \ \mathrm{cells/s}= 81 \ \mathrm{km/h}$ and a length of two cells. 

Having two vehicles $i$ and $p$ (being the latter in front of the former) on the same lane, with positions $x_i$, $x_p$ and speeds $v_i$, $v_p$ respectively, then the distance or empty cells between them is defined as $d_i := x_p-x_i-l_p$. Where $l_p$ is the length of vehicle $p$. The positions $x_i$ and $x_p$ are considered as the cell number. So, if the vehicle $p$ is in cell $n$, then $x_p=n$. In the case of slow vehicles, their position is considered to be as the front cell. \\

Wolfram studied Cellular Automata during the second half of the twentieth century and for the first decade of this century \cite{Wolfram}. The goal was to create a framework where a model based in discrete space and time could create complex (in the most basic sense of the word) behaviors with simple rules. In that sense, each discrete part of the space, or ``cell'', follows the exact same rules as the others with the exact same order. An interesting feature of Cellular Automata (then defined by the space and the set of rules) is their scalability. ``Complexity'' can be escalated not only by the number of rules but also with how the rules make cells interact between themselves.   

Modern computational tools allow implementing a set of rules in two different ways: parallel and serial. The former refers to parallel computation using GPUs and CPUs. Serial implementation, as the names suggest, refers to the execution of tasks by an only set of processors in a specific order. Nagel and Schreckenberg \cite{NaSch} follow a serial order.

As said before, the highway studied here is composed by a discrete space. Only the right set of rules are needed in order to model a proper behavior of the cars in it. In order to do so, Nagel and Schreckenberg \cite{NaSch} decomposed how a driver-car works in four different actions. Intuitively, drivers want to go as fast as possible. However, they are limited by different factors like a maximum speed limit, other cars, or random incidents (sun glare, people crossing, etc.). Drivers thus need to find how much they can accelerate taking into account all these elements. Nonetheless, other elements are involved here. There is not one unique kind of driver, but a whole spectrum. A set of different driving styles is allowed, but only one is chosen and fixed. Car crashes are not allowed. The decomposition of the driver-car and is thus put into four rules as in \cite{NaSch,larragadelrio2005}.

\begin{description}
\item[R1 -- Acceleration] If $v_i < v_{\mathrm{max}}$, the speed of the car $i$ is increased by one unit.
\begin{align}
v_i \gets \mathrm{min}(v_i+1, v_{\mathrm{max}}).
\end{align}

\item[R2 -- Randomization] If $v_i>0$, the speed of the car $i$ is randomly decreased by one unit with probability $R$.
\begin{align}
v_i \gets \mathrm{max}(v_i-1, 0) \ \mathrm{with \ probability} \ R.
\end{align}

\item[R3 -- Deceleration] If $v_i> d_i^s$, with $d_i^s = d_i + \lceil (1-\alpha)v_p\rceil,$
where $\alpha \in [0, 1]$ and $\lceil x \rceil$ denotes the smallest following integer from $x$, then the speed of vehicle $i$ is decreased.
\begin{align}
v_i \gets \mathrm{min}(v_i, d_i^s).
\end{align}

\item[R4 -- Movement] The vehicle $i$ is moved forward with the new speed computed by R1-3.
\begin{align}
x_i \gets x_i+v_i.
\end{align}
\end{description}

The first three rules modify the speed independently of the position. The fourth rule gives the new position of the vehicle $i$ considering the new speed. 
R1 and R3 assure that all vehicles will go to the maximum speed available and will slow down when needed to avoid a crash. The second rule has a stochastic parameter to model random deceleration while driving. 
In comparison with the NaSch model \cite{NaSch}, we commute R2 and R3 to avoid accidents.

The deceleration rule (R3) involves a parameter $\alpha$ to model different types of driving. $\alpha$ is called \textit{anticipatory driving parameter} set. When $\alpha = 1$ the speed of the vehicle ahead is not considered. This case can be compared to a very aggressive style of driving when the vehicle behind will be very close to the one ahead. Getting closer to $0$ then means a very cautious way of driving, leaving big distances between one car and another. 
 
 \subsection*{\label{ssec:cambios}Merging into other lanes}
 
When expanding the model to several lanes, each one of them follows the same set of rules. The possibility from the vehicles to merge into other lanes is available with considerations.

Mexico's laws prohibit a vehicle to pass another one by the right lane. This creates a distinction of lanes, being the extreme left lane as the one where vehicles with greater speed drive and the extreme right one where slow drivers are found.

In that sense, a prohibition to right-pass for all types of vehicle is imposed. Also, slow vehicles are limited to be in the left lane only when passing is needed. In the case where the highway has three lanes, slow vehicles are limited to drive on the middle and right lane only.
Thus, the conditions for every vehicle to merge are as follows: 

\begin{description}
	\item[Incentive] The vehicle ahead must go slower.
    \begin{gather}
		d_i^s < v_i.
	\end{gather}

	\item[Safety 1] The vehicle behind on the objective lane must go slow enough to avoid a crash. 
    \begin{align}
		d_b^{s} > v_b.
	\end{align}
    
	\item[Safety 2] The vehicle ahead on the objective lane must be far enough to avoid a crash.
   \begin{align}
		d_i^{s'} > v_i.
	\end{align} 
 \end{description} 

When merging into a lefter lane the three conditions are required for both slow and fast vehicles. When merging into a righter lane, the distinction between each type of vehicle lays on the urgency of slow vehicles to do it. As a measure for assuring this, the \textbf{Incentive} is only applied on fast vehicles wanting to merge onto a righter lane.

 \subsection*{\label{ssec:ramps}Ramps}

\begin{figure}[ht]
\centering
\includegraphics[width=0.6\textwidth]{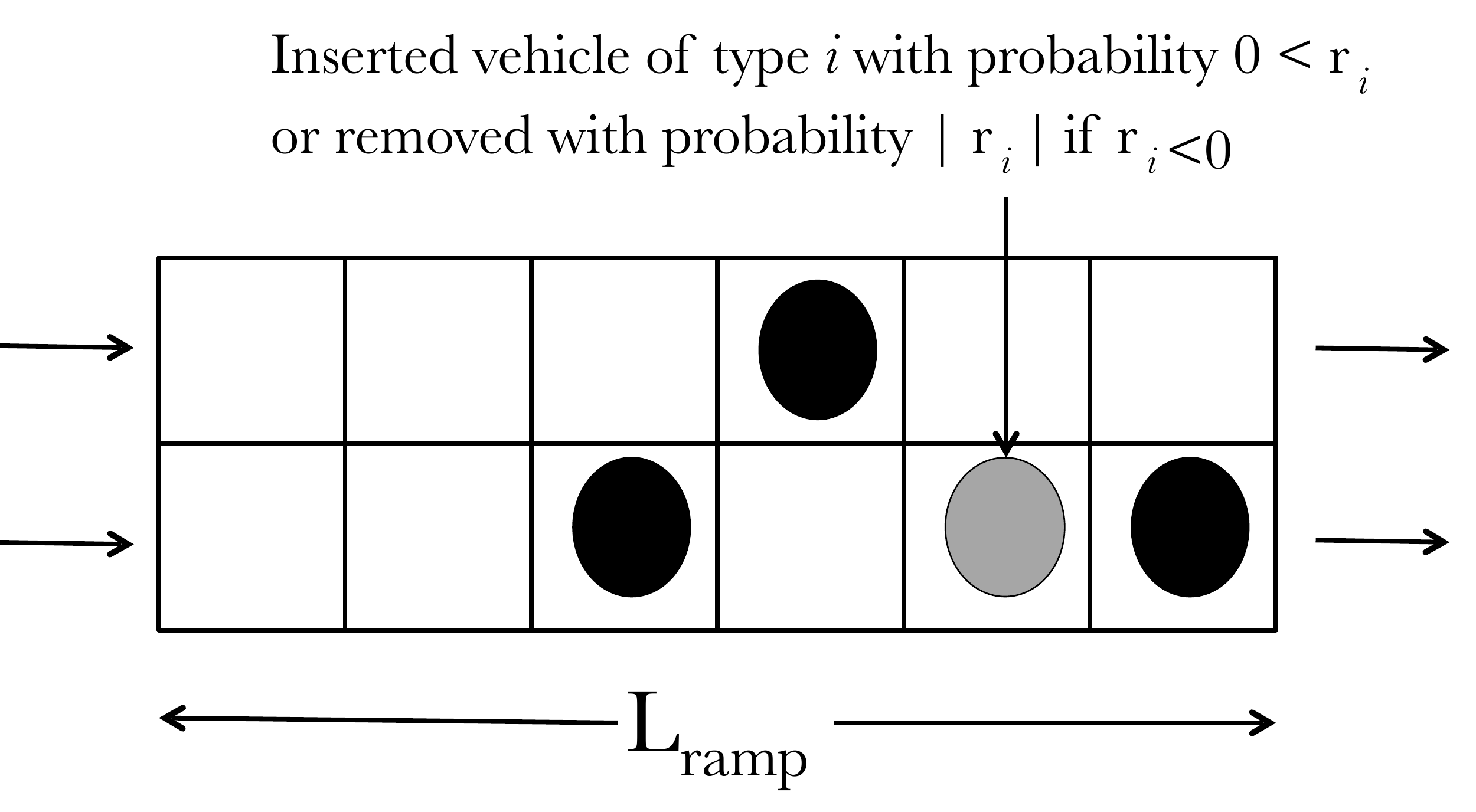}
\caption{\label{fig:ramp}Scheme of a ramp in the right lane, with a length of $L_\mathrm{ramp}$. A vehicle is inserted into the ramp with rate $|r_\mathrm{init}$| of type 1 with probability $r_1$ or type 2 with probability $1-r_1$.}
\end{figure}

A ``physical'' ramp is not integrated \cite{LarragaDoc}, but we rather model the capacity of a vehicle to enter/quit the system on a section of the highway. To each ramp two different rates $-1\leq r_1, r_2\leq 1$ are associated, so at each time-step a vehicle of type 1 or 2 can enter or quit the system. 
In Figure \ref{fig:ramp}, a scheme of the ramp is observed.

When $r_i>0$, a vehicle of type $i$ with speed $v = 2 \ \mathrm{cells/s} = 54 \ \mathrm{km/h}$ is inserted into the system before the four rules R1-R4 are applied with probability $r_i$. The vehicle is inserted without consideration of safety distance, meaning it will be put into the first empty cell found, without considering its neighbors. 
When $r_i<0$, then a vehicle of type $i$ in the ramp zone is deleted from the system with probability $|r_i|$.

In that sense, at each time-step, every ramp might introduce or remove two vehicles of different types, or simply introduce one of any type.

To model the initial flux, a third rate $0\leq r_\mathrm{init}\leq 1$ is introduced so there are a maximum of $r_\mathrm{init}\times 3600$ cars entering to the system per hour per lane. The type of vehicle inserted at the beginning of the highway is given by the respective rate $r_1$. Doing a small summary for the presented rates, for a sequence of random tests:
\begin{itemize}
\item An initial fast vehicle is inserted if \texttt{rand()} $\leq r_\mathrm{init}$ and \texttt{rand()} $> r_1$;
\item An initial slow vehicle is inserted if \texttt{rand()} $\leq r_\mathrm{init}$ and \texttt{rand()} $\leq r_1$;
\item A fast vehicle is inserted in a ramp if \texttt{rand()} $\leq r_2$ with $r_2>0$;
\item A slow vehicle is inserted in a ramp if \texttt{rand()} $\leq r_1$ with $r_1>0$;
\item A fast vehicle exits at a ramp if \texttt{rand()} $\leq |r_2|$ with $r_2<0$;
\item A slow vehicle exits at a ramp if \texttt{rand()} $\leq |r_1|$ with $r_1<0$.
\end{itemize}

\section{\label{ssec:Data}Available data}

The Mexican Secretariat (Ministry) of Communications and Transport publishes each year a table with data from most of the federal highways and roads\footnote{http://www.sct.gob.mx/carreteras/direccion-general-de-servicios-tecnicos/datos-viales/}. Data presents the Annual Average Daily Traffic (AADT) and the composition of that measurement as a proportion of motorcycles, small vehicles and large vehicles for a given measurement station with a given location and a given direction. AADT is the total volume of vehicle traffic of a highway or road in a year, divided by 365 days. Measurements are taken in three different types of reported stations: type 1 where flux is measured before the ``traffic generating point'', type 2 measuring at the generating point, and type 3 measuring after the generating point.

\begin{figure}
\begin{subfigure}{0.49\textwidth}
\includegraphics[width=1.\textwidth]{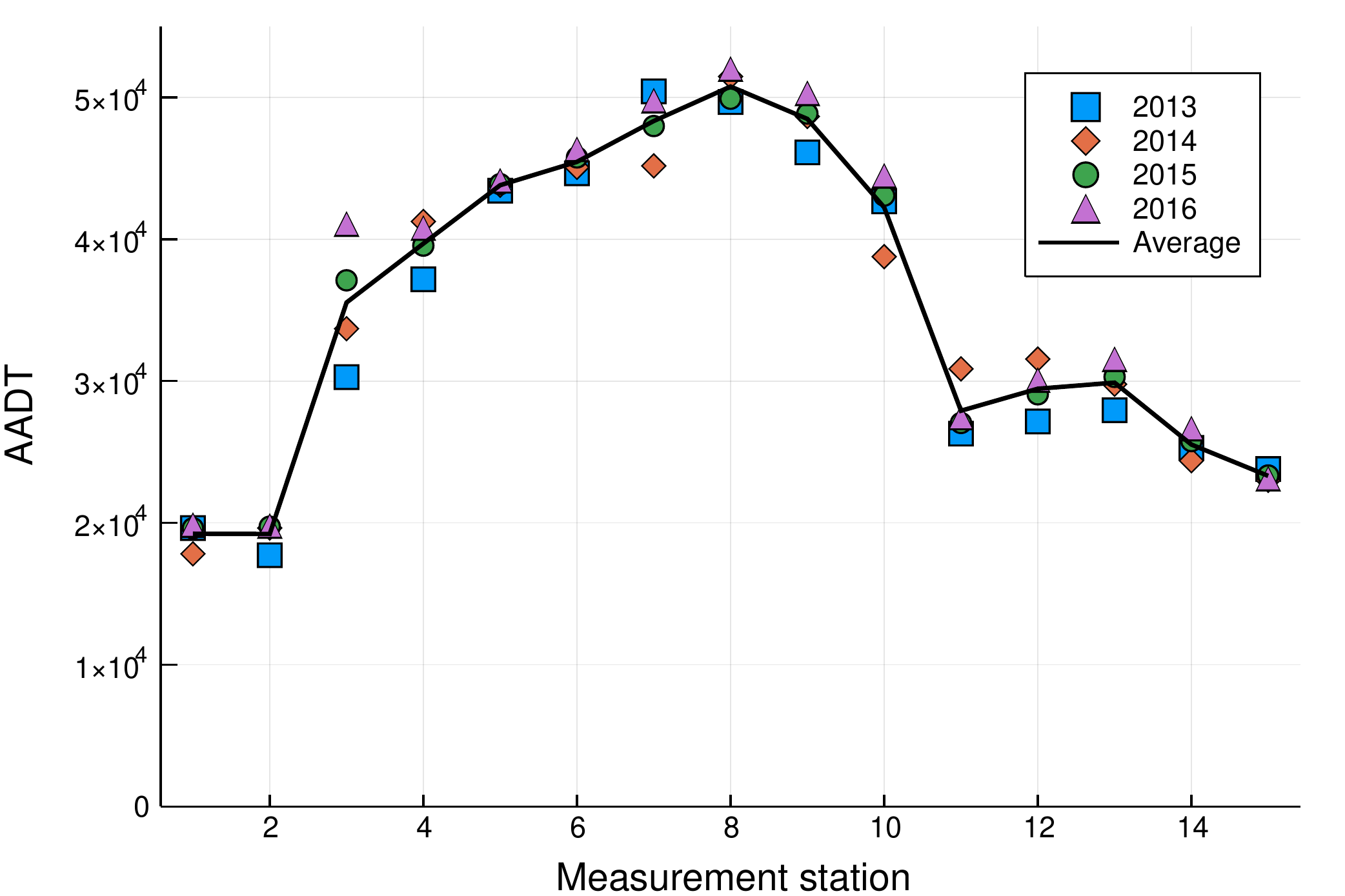}
\caption{\label{fig:S1-AADT}}
\end{subfigure}
\hfill
\begin{subfigure}{0.49\textwidth}
\includegraphics[width=1.\textwidth]{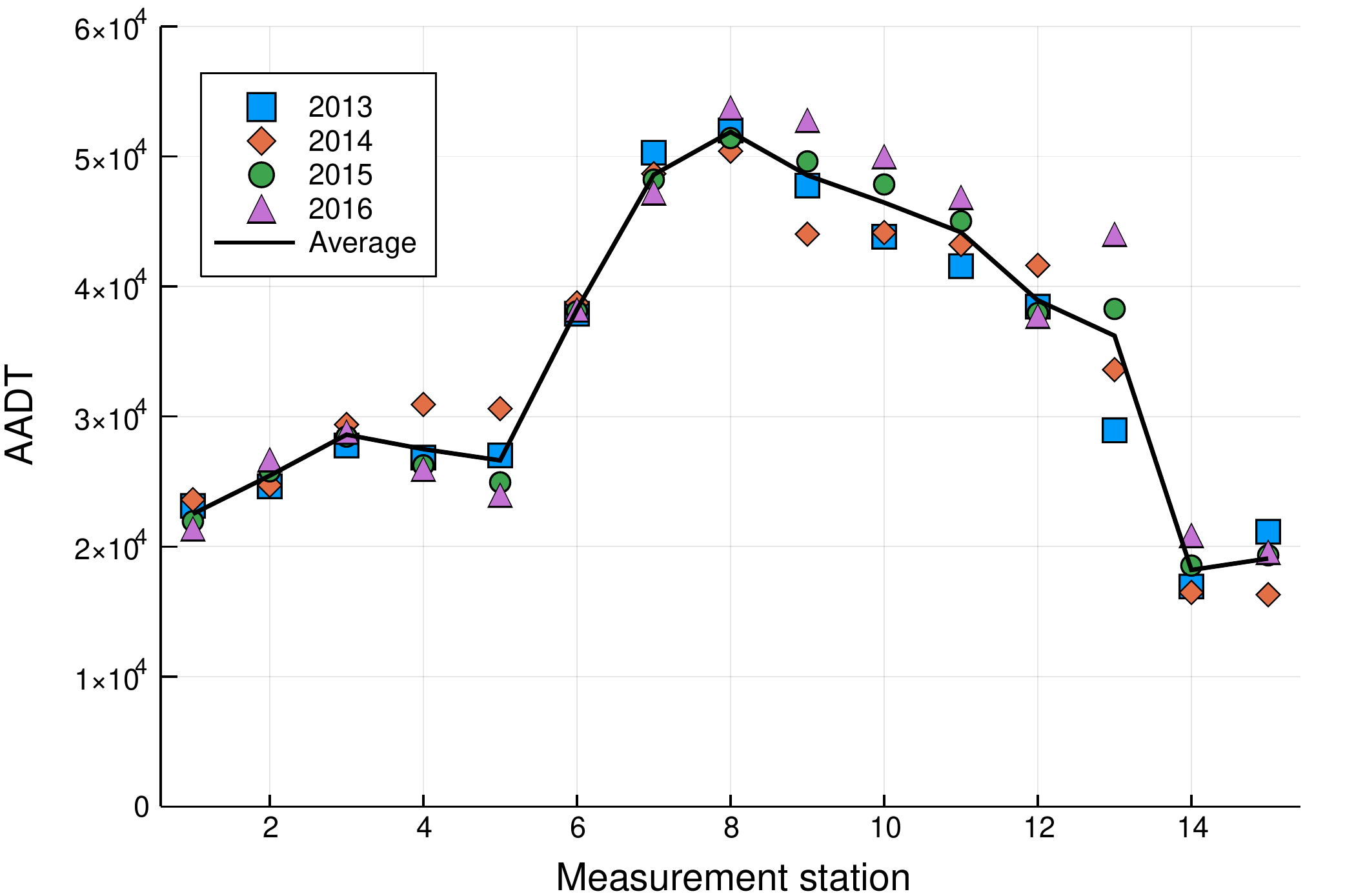}
\caption{\label{fig:S2-AADT}}
\end{subfigure}
\hfill
\begin{subfigure}{0.49\textwidth}
\includegraphics[width=1.\textwidth]{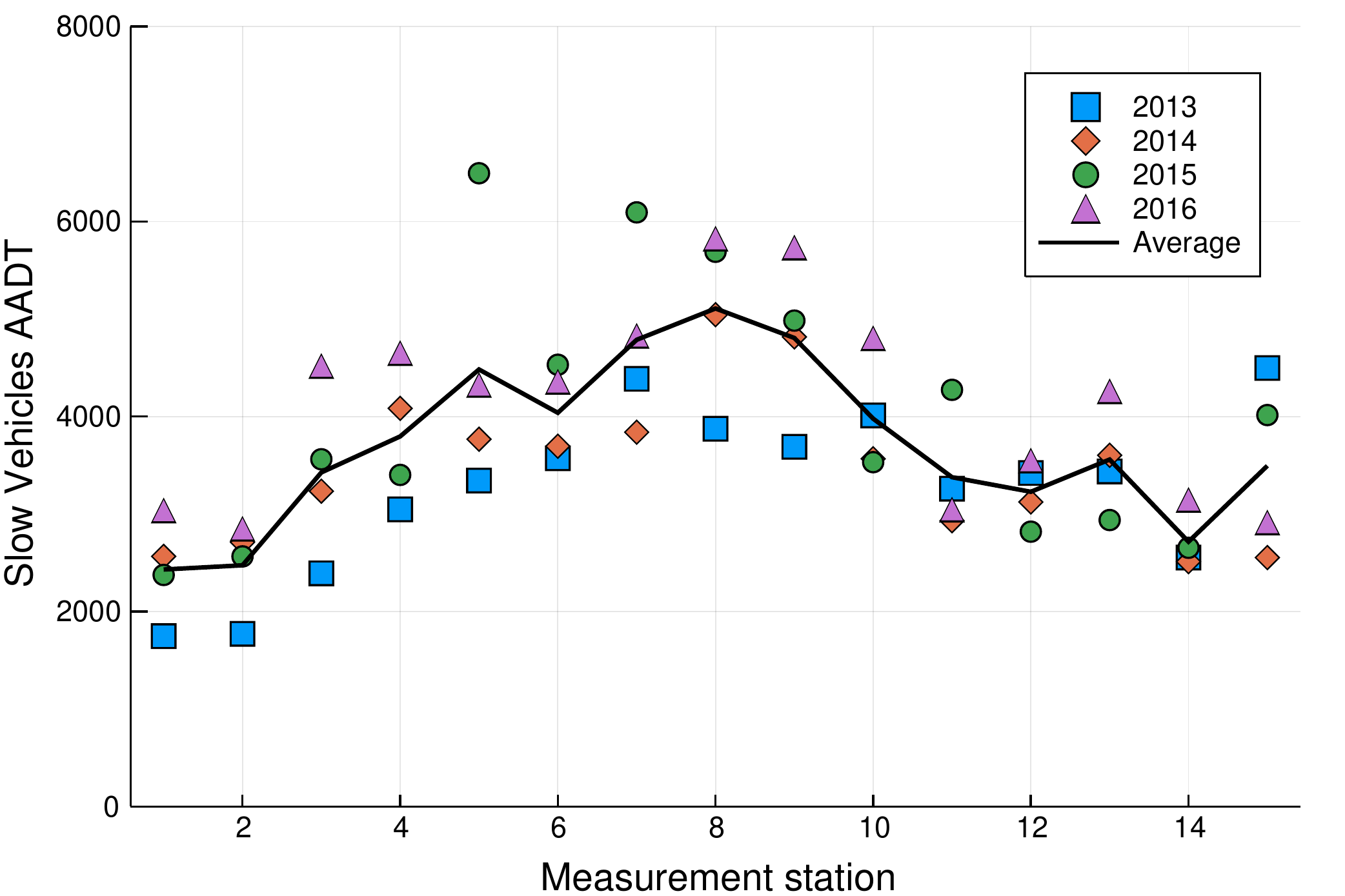}
\caption{\label{fig:S1-slow}}
\end{subfigure}
\begin{subfigure}{0.49\textwidth}
\includegraphics[width=1.\textwidth]{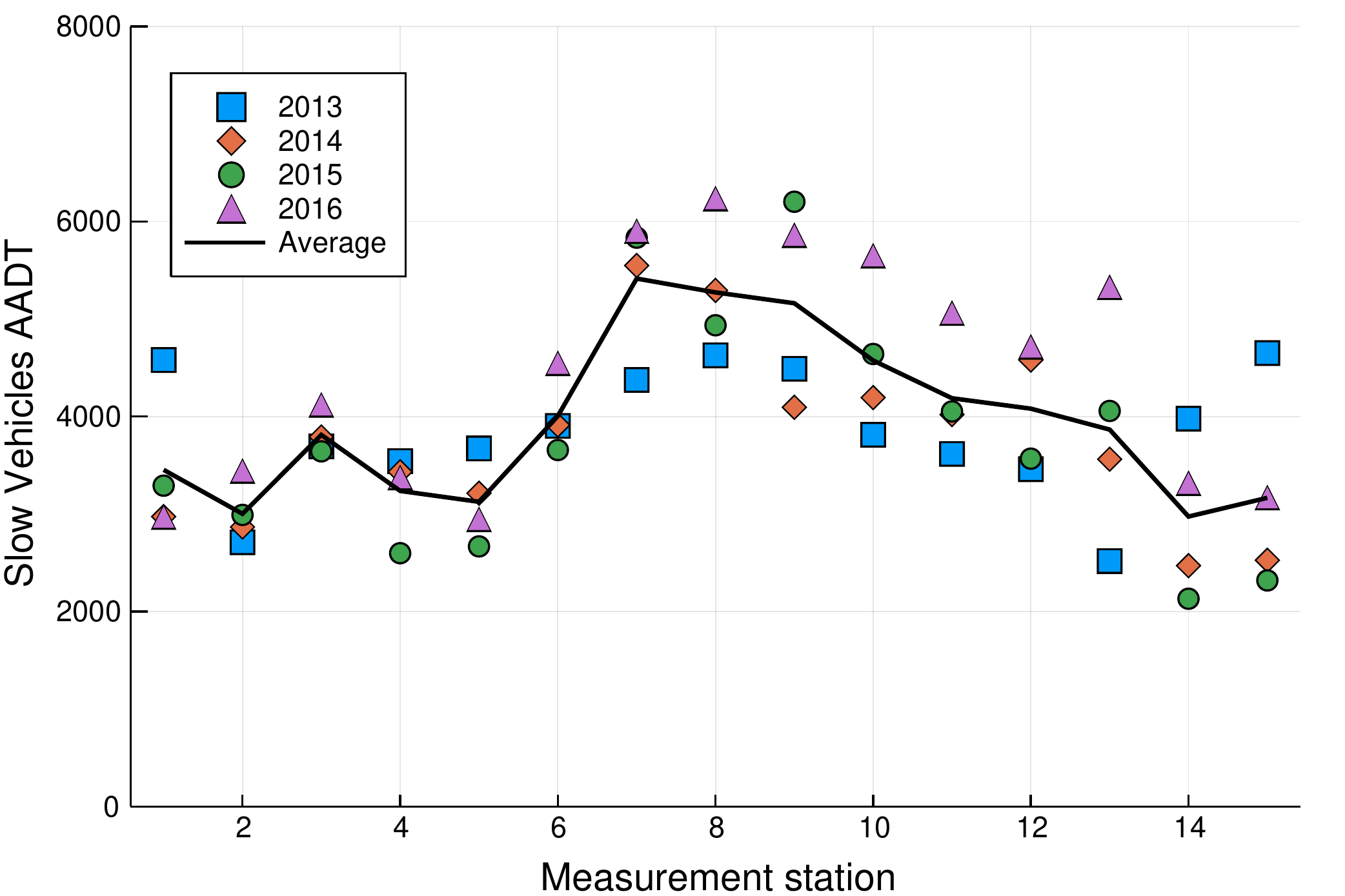}
\caption{\label{fig:S2-slow}}
\end{subfigure}

\caption{\label{fig:data}Annual average daily traffic of cars per year for S1 and S2. Measurement stations are listed as driving through each sense. Odd measurement stations are type 1, whereas even stations are type 3. (a): All kind of vehicles, S1. (b): All kind of vehicles, S2. (c): Only slow vehicles, S1. (d): Only slow vehicles, S2.}
\end{figure}

In the case of the Cuernavaca Bypass, data from 2013 to 2016 is taken. Before this period, there is missing information of the 27.3 km studied, while in the four years taken the information is consistent in the way and the place where flux is measured. Figure \ref{fig:data} presents the total and slow vehicles AADT data from the four years for S1 and S2 and the average of this raw data (black bold line).

Data is processed to obtain the net flux per hour in the different ramps of both senses taking the temporal average of the obtained data. First, we suppose that there is an homogeneity of flux during 12 hours per day. Given the location of a ramp, the measurements of the immediate station after (type 3) and before the ramp (type 1) are taken. The difference is computed and then divided by 12 hours.
Supposing that the flux in the ramps is homogeneous during 12 hours per day the 7 days of the week is a poor approximation. However, the information obtained from the official data is insufficient to make a more accurate analysis given the number of ramps and the different dynamics they can have during the different hours and days of the week.

\begin{figure}
\begin{subfigure}{0.49\textwidth}
\includegraphics[width=1.\textwidth]{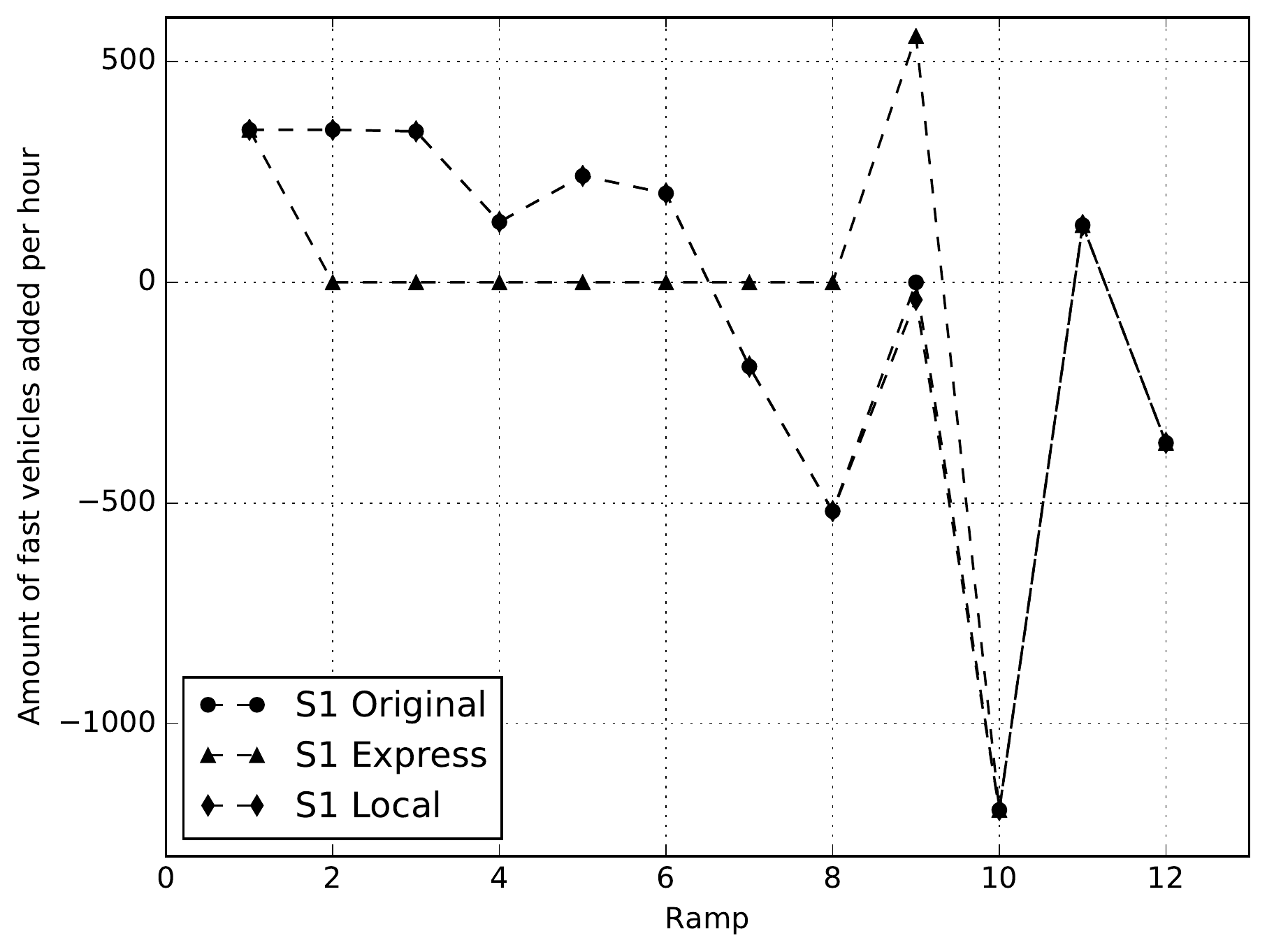}
\caption{\label{fig:rampsS1-2}}
\end{subfigure}
\hfill
\begin{subfigure}{0.49\textwidth}
\includegraphics[width=1.\textwidth]{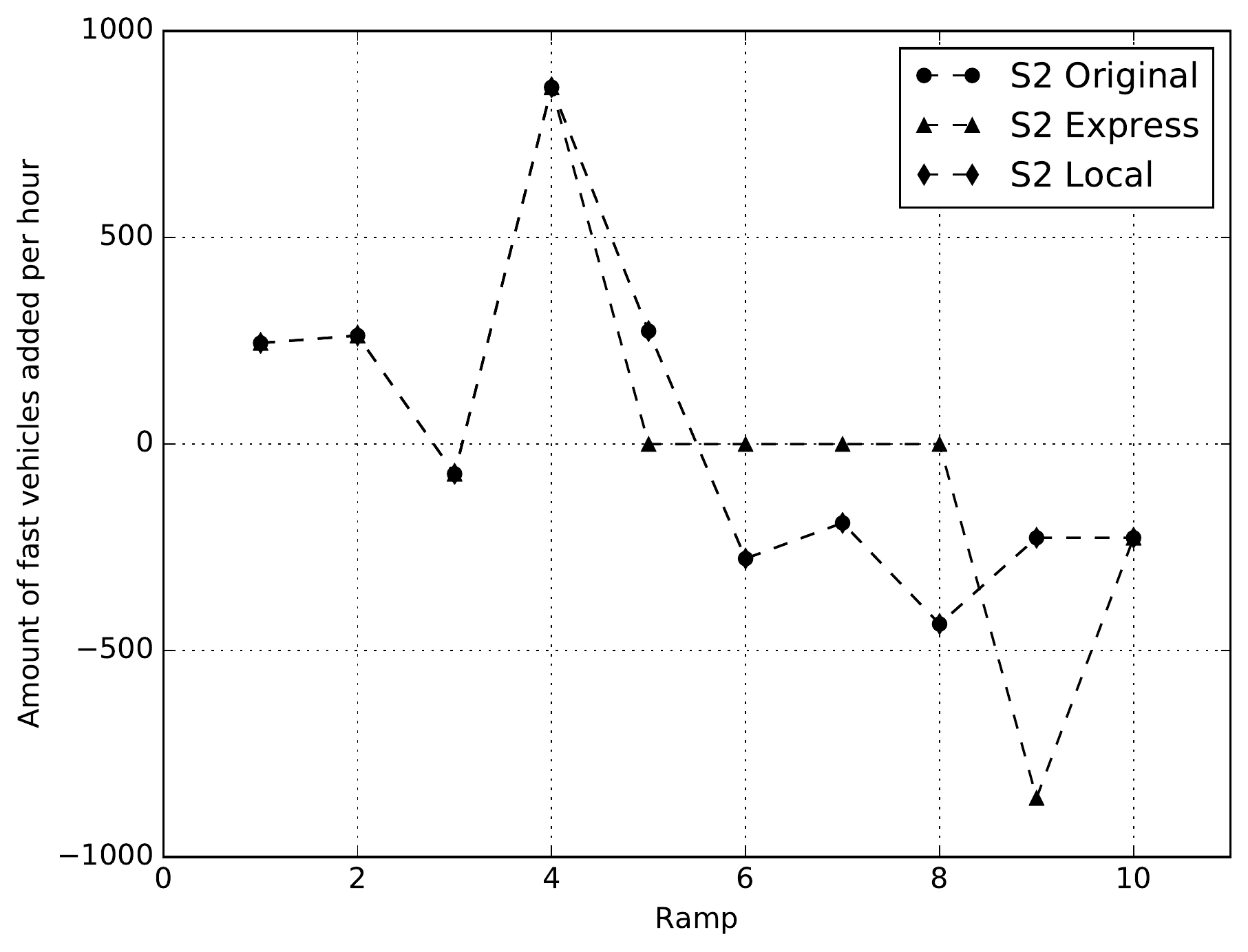}
\caption{\label{fig:rampsS2-2}}
\end{subfigure}
\hfill
\begin{subfigure}{0.49\textwidth}
\includegraphics[width=1.\textwidth]{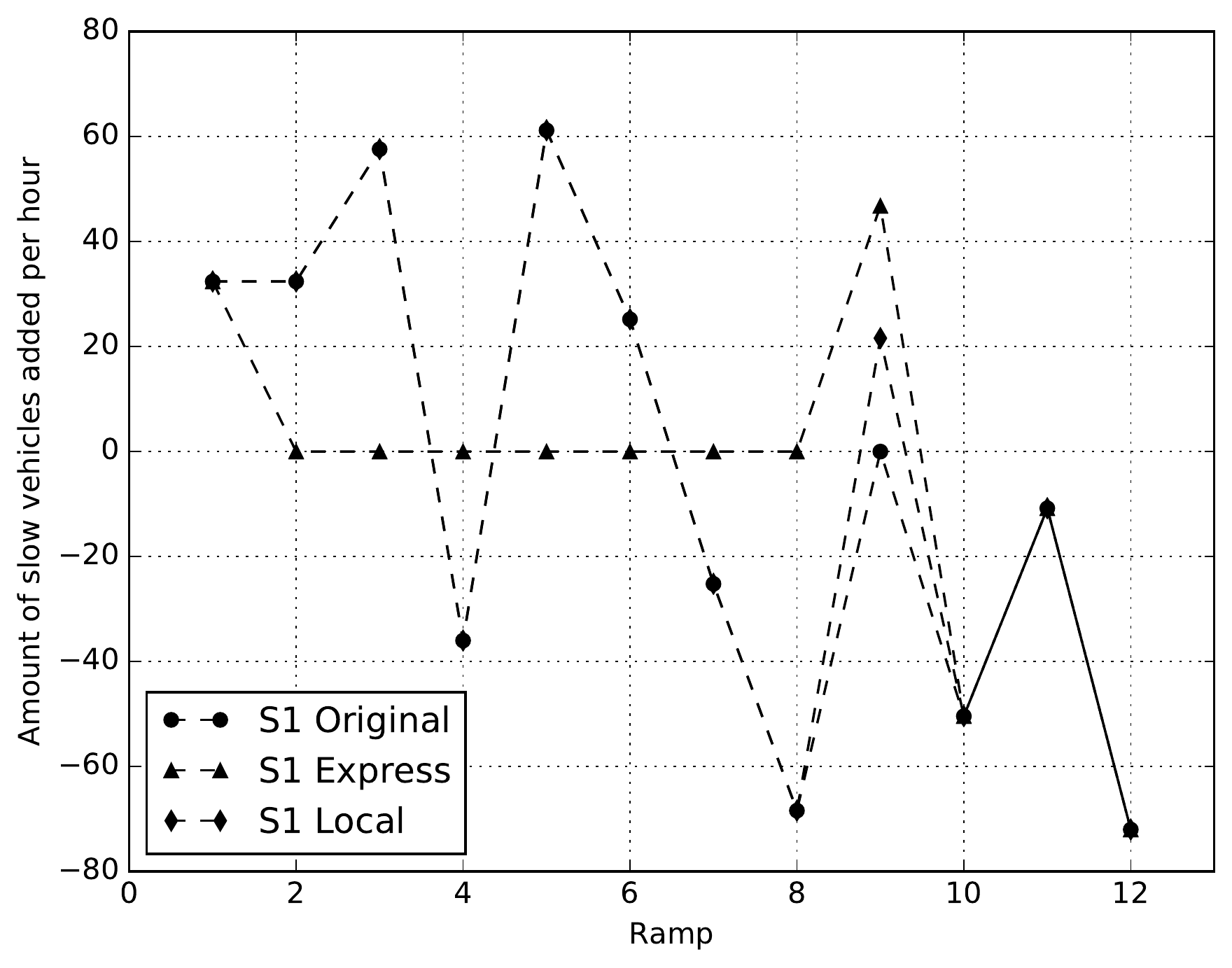}
\caption{\label{fig:rampsS1-1}}
\end{subfigure}
\begin{subfigure}{0.49\textwidth}
\includegraphics[width=1.\textwidth]{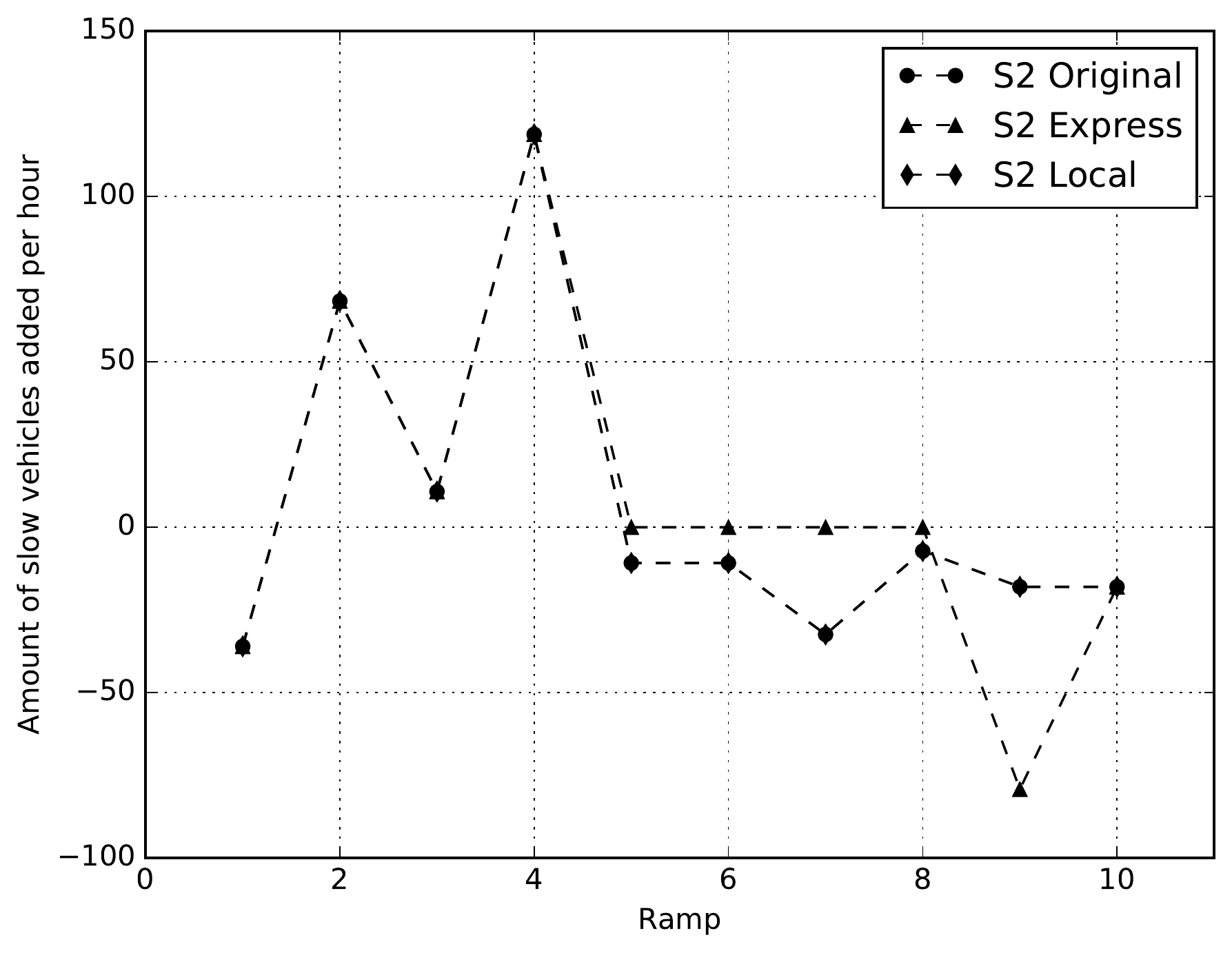}
\caption{\label{fig:rampsS2-1}}
\end{subfigure}

\caption{\label{fig:ramps}Flow added at each ramp of the six studied systems. (a) and (b) refer to the fast vehicles flux at the ramps of S1 and S2 respectively. (c) and (d) refer to slow vehicles, also for S1 and S2 respectively. Points showing 0 vehicles per hour mean that the ramp was removed.}
\end{figure}

We call, for matters of simplification, $\bar{f}_r$ the net flux in ramp $r$ per hour. That is, $\bar{f}=(\mathrm{AADT}_{r, 3}-\mathrm{AADT}_{r,1})/12$, where AADT$_{r, i}$ is the documented AADT at station of type $i$ of ramp $r$. To obtain a rate $r_i$ as in Section \ref{ssec:ramps}, then we divide $\bar{f}_r$ by 3600. The results of this computation is presented in Figure \ref{fig:ramps}.

\section{\label{sec:Methods}Methods}

\subsection{\label{ssec:phases} Phases in traffic theory}
Generally, a phase is defined as a state in space and time.  This concept has been initially used in areas such as physics,  chemistry, and thermodynamics. In these systems, “phases” mean different aggregate states (such as solid, fluid, or gaseous; or different material compositions in metallurgy; or different collective states in solid state physics) \cite{treiber}.  Most of the time,  the terms phase and phase diagram are applied to large (quasi-infinite), spatially closed, and homogeneous systems in thermodynamic equilibrium, where the phase can be determined in any point of the system.  However, when these concepts are transferred to traffic flows, researchers have distinguished between one-phase, two-phases, and three-phases models.  In particular, the number of phases is mainly related to the number of traffic flow states that the instability diagram distinguishes.

Classical theories based on the fundamental diagram of traffic flow have two phases: free flow and congested traffic. Kerner \cite{kerner1, kerner2, kerner3} developed a theory describing three phases, adding a synchronized phase. However, Kerner defined the synchronized phase as a unperturbed flow with decreasing flow. In this case we follow other definition of synchronized flow where an unperturbed mean speed is observed \cite{Schreckenberg2015}. Nonetheless, from Kerner we recover the names he used to describe the different phases. These are:
Free flow (F), Synchronized flow (S), Wide moving jam (J). In free flow, vehicles travel at a maximum speed, which is dependent, amongst other things, on the design speed of a road, the weather and the speed restrictions in operation at any particular time. After the congestion transition occurs, the free flow changes to synchronized flow. In the synchronized flow, the speed of vehicles drops significantly from the maximum speed. When the synchronization phase passes, jams start to occur, so one can find vehicles with null speed. At this phase, the greater is the density, the less will be the flow, up to the point where no car will be able to move. At this maximum density, the flow will be zero again.

\begin{figure}
    \centering
    \includegraphics[width=0.6\textwidth]{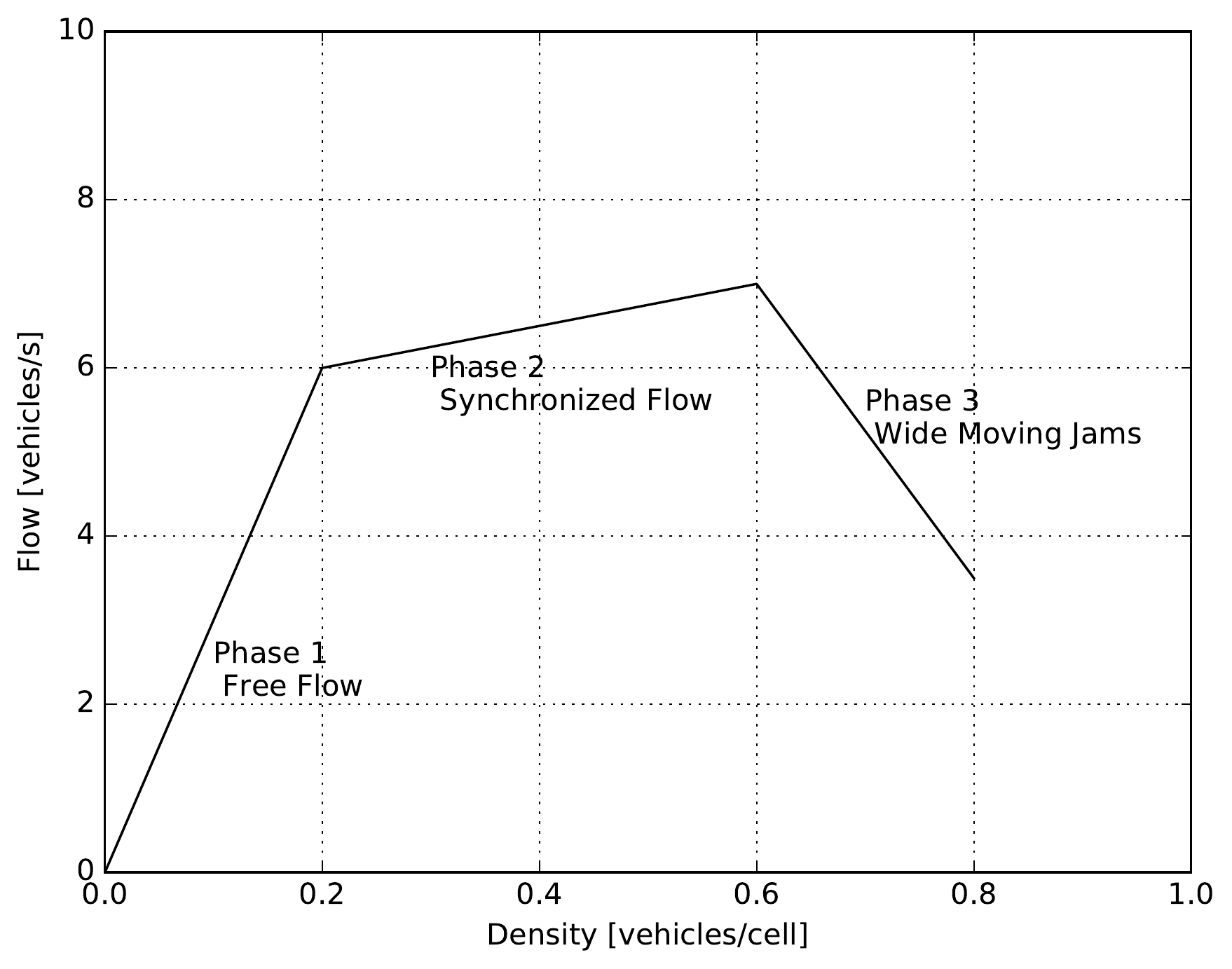}
    \caption{\label{fig:fundamental}Transition phases in a fundamental diagram (see ref. \cite{Schad1}).}
\end{figure}

In Figure \ref{fig:fundamental}, a schematic of the three described phases is presented. Phase 1 is the non-congested phase (free flow) when there is no influence of the increasing density on the speeds of the vehicles. The speed does not drop with the introduction of new vehicles into the freeway. Here we can stress that the change between free flow and synchronized flow may occur at higher densities for automated vehicles according with anticipatory behavior \cite{delRio}, \cite{Schad1}  Phase 2 finds the freeway cannot sustain the slope with the injection of newer vehicles into the traffic stream. In phase 3, the system enters into a congested state where the flow starts to decrease as the density increases.

\subsection{\label{ssec:physics}Physical elements for vehicular traffic analysis and its limitations}

Studying the macroscopic variables obtained from a highway such as a local density and a local flux, an organic nature to apply analysis tools from fluids theory may come. However, these tools must be applied with subtle care. More specifically, in this section, we will refer to the fundamental diagram usually used to detect and examine phase transitions in physics.

As a reminder, fundamental diagrams are plots of flux vs. density. In a slow density regime, flux increases linearly as density increases. It is in this regime where the fundamental relation of vehicular transit $j = \rho v$ comes, $j$ being the flux, $\rho$ the density, and $v$ the speed. As described in the Section \ref{ssec:phases}, we are talking about a free flow regime. However, in order for this relationship to be met, the way to measure is important and can determine the information in a fundamental diagram\cite{Maerivoet}.

In normal conditions, when the flux ceases to increase proportional to the maximum speed as the density increases, the free flux regime finishes. More specifically, the regime finishes when flux meets a maximum concerning density. This maximum is known as the \textit{capacity} of the system. Once this point is passed, the system enters into a \textit{congested state}. 

However, it is at this point where things start to complicate. In a first place, it is imperative to know what a measurement plotted in a fundamental diagram is. This kind of diagram is thought to present one-dimensional equilibrium curves. In order to achieve that, equilibrium must be met. Nevertheless, in traffic analysis, equilibrium can be met only if the system is \textit{closed} and \textit{unperturbed}. Translating into a more physical language, walls must be adiabatic, i.e. no ramps. In an open system with ramps, such as the one studied here, equilibrium cannot be met, but a stationary state can. This is only because the flux entering at the starting point of the highway and the ramps are set as macroscopically constant over time. Thus, fundamental diagrams cannot be used in the present study to detect phase transitions, and other tools must be used to.

Also, vehicular systems might present different phenomena between a free flux regime and a congested state, e.g. hysteresis\cite{Maerivoet} or what is known as a synchronization period\cite{schadschneider2000statistical}. Both might be represented in a fundamental diagram (figures in references \cite{Maerivoet,schadschneider2000statistical}) and have a translation to physical systems. Hysteresis is evident and will not be developed. A synchronization period, however, might be seen in an analogy with states of the matter as a mixed phase between gas and liquid. In this phase, vehicles might find a free path where they can go as quick as they want, but they can also be enclosed within a group with a uniform speed. The speed of these groups is determined by the group-leader which in our case can be a slow vehicle or a fast vehicle. 

In that sense, the gas-liquid analogy can be used to distinguish phases.  A free flux regime is called \textit{gaseous state} where vehicles might go to a mean speed close to its maximum value; a first synchronized state called \textit{light mixed state} where only fast vehicles led-groups can be found, and a second synchronized state called \textit{heavy mixed state} where slow vehicles led-groups can also be found. Finally, a \textit{liquid state} equivalent to a full congested state. The analogy can continue up to a ``solid state'' where the highway is fully stopped, and no car moves. However, such a scenario is not present in this study.

\subsection{\label{ssec:stdev}Standard deviation as a measure of order}

The different phases in a vehicular system reveal how the relation of the macroscopic variables such as speed, flux, and density might change. However, looking closer into the microscopical level of the highway, a phase transition can be related to the apparition of specific structures created by the vehicles moving (or not), such as the led-groups described in Section \ref{ssec:physics}.

Traditionally, entropy is a good choice to measure order. Nevertheless, the problem comes to how to measure it and to give a solid definition in our case. Indeed it has already been discussed how entropy can be omitted to ``measure'' order \cite{Reiss1986}. In addition to this, being a macroscopic variable, an entropy measurement does not give information about the formation of microscopic structures in the highway. 

As an alternative to the entropy, the analysis of the standard deviation of the macroscopic variables is done \cite{Schad1,schadschneider2000statistical, Reiss1986}. Phase transitions are easily detected doing this as it is further shown in Section \ref{ssec:Unperturbed}. The intuitive idea behind this is the following: imagine a highway scarcely driven. If a first vehicle diminishes its speed by a random reason (noise), it is quite probable that the vehicle behind will not be affected because of the large distance between both of them. As the highway is filled, the distance between two cars diminishes, resulting in the fact that a change in the speed of the first vehicle will more probably affect the next behind, and the next behind, and the next behind, etc.
If the mean speed and its standard deviation of both scenarios are measured, it is found that the scarcely driven system has a bigger standard deviation than the more dense one. In this sense, standard deviation gives information about the order in the system and is also related to the probability of having a crash with another car\cite{delRio,Schad1}.  

However, what happens in a phase transition? Imagine a highway where the cars follow a free flow regime. As said before, the standard deviation decreases as the density increases. If a phase transition comes into a synchronization phase, then there will be an interval of density where the whole systems rearrange itself in order to change its microstructure into a new phase. This complete rearrangement means that many vehicles need to change their speed (possibly violently), so the new structure can arise. In this sense, during a phase transition, a significant change in the speed standard deviation must be observed. Once the system is in a new phase, the order is restored, and the standard deviation can decrease again.
Thus, phase transitions are related to peaks in the speed standard deviation.

\subsection{\label{ssec:CO2}Computing Travel Times}

Travel times are useful to compute as it gives a more intuitive idea and easy-to-rely impression of what is happening. Highways in this kind of models are used to be divided in different sections, where each end of a section can be delimited by ramps or measurement stations, or any other real or fictional limit. In this case of study, sections are delimited by ramps to assure that the flux in a given section remains unperturbed, thus resulting in a steady state.

Knowing the average speed of a given type of vehicle for each section of a lane $\ell$, $\langle v\rangle_{s, \ell}$ and the corresponding length of the section, a simple division between these two quantities gives the average time to cross that section $T_{s, \ell}$. Knowing the average use of the specific lane in that particular section $u_{s, \ell}=j_{s, \ell}/\sum_\ell j_{s, \ell}$ will finally give the average time to cross all sections, 
\begin{equation}
\label{eq:Times}
T_t = \sum_s\sum_\ell u_{s, \ell}T_{s, \ell}= \sum_s\sum_\ell u_{s, \ell}\frac{\mathrm{Length}_\ell}{\langle v\rangle_{s, \ell}} .
\end{equation}
In this case, the use $u_{s, \ell}$ refers to the portion of vehicles of a given type that are in lane $\ell$. Thus, $\sum_\ell u_{s,\ell}=1$.

\section{\label{sec:Results}Results}

The bypass is divided into different sections with lengths varying from 150 to 250 meters. S1 is divided into 18 sections and S2 in 19. Measurements are done at the end of each section by a unique sensor measuring individual speeds and fluxes during a period of 12 seconds and then having an average temporal speed over the same period. The proportion of slow vehicles $r_1$ is swept from 0\% to 40\% taking steps of 2\%. The initial flux $r_\mathrm{init}$ is also swept increasing three vehicles per minute per lane from 0 veh/(h lane) to 2340 veh/(h lane). Simulations are not transitory, meaning that the initial flux and the proportion of slow vehicles remain constant during the whole computation time. An ensemble of $N =$ 101 of simulations is taken for each combination of ($r_1, r_\mathrm{init}$). An ``average system'' over the ensemble is done to have a unique temporal system. It is this final system which is analyzed over time. It is important to note that results do not permute because of the lack of equilibrium and the stationary state of the system. So, to do a first temporal average of any simulation and then a study over the ensemble gives different results.

Parameters used are $\alpha=0.75$, $R=0.2$ \cite{delRio}. In order to meet the topographical conditions imposed on the maximum speed of each type of vehicle discussed in Section \ref{sec:Highway}, the top speeds imposed are $v_\mathrm{max, fast}=4$ cells/s = 108 km/h in dangerous curves, and $v_\mathrm{max, slow}=2$ cells/s = 54 km/h in the slope.

In a first moment, the study of the unperturbed system is done. The unperturbed system means that every ramp is taken off and only the initial flux, the proportion of slow vehicles and the topographical elements of the highway play a role. Then, ramps are inserted again to study the Cuernavaca bypass and its modifications in Section \ref{ssec:Perturbed}. The ramps have the rates presented in Section \ref{ssec:Data}.

\subsection{\label{ssec:Unperturbed}Unperturbed system}

The study of the unperturbed system is done with two different purposes. First, to show and explain the concepts discussed in Section \ref{ssec:stdev}. Second, to study the effects of the topographical elements in the system. As said, only the initial flux, the proportion of slow vehicles and the topographical elements of both senses influence the results. In that sense, S1 is omitted in this section as the dangerous curves (topographical elements of S1) are in the first 5 kilometers of the highway, so there is no important feature to report. On the other hand, S2, because of the presence of both curves and the slope affecting fast and slow vehicles respectively in the second half of the highway makes it more interesting to study and report. 

\begin{figure}[ht]
\begin{subfigure}{0.49\textwidth}
\includegraphics[width=\textwidth]{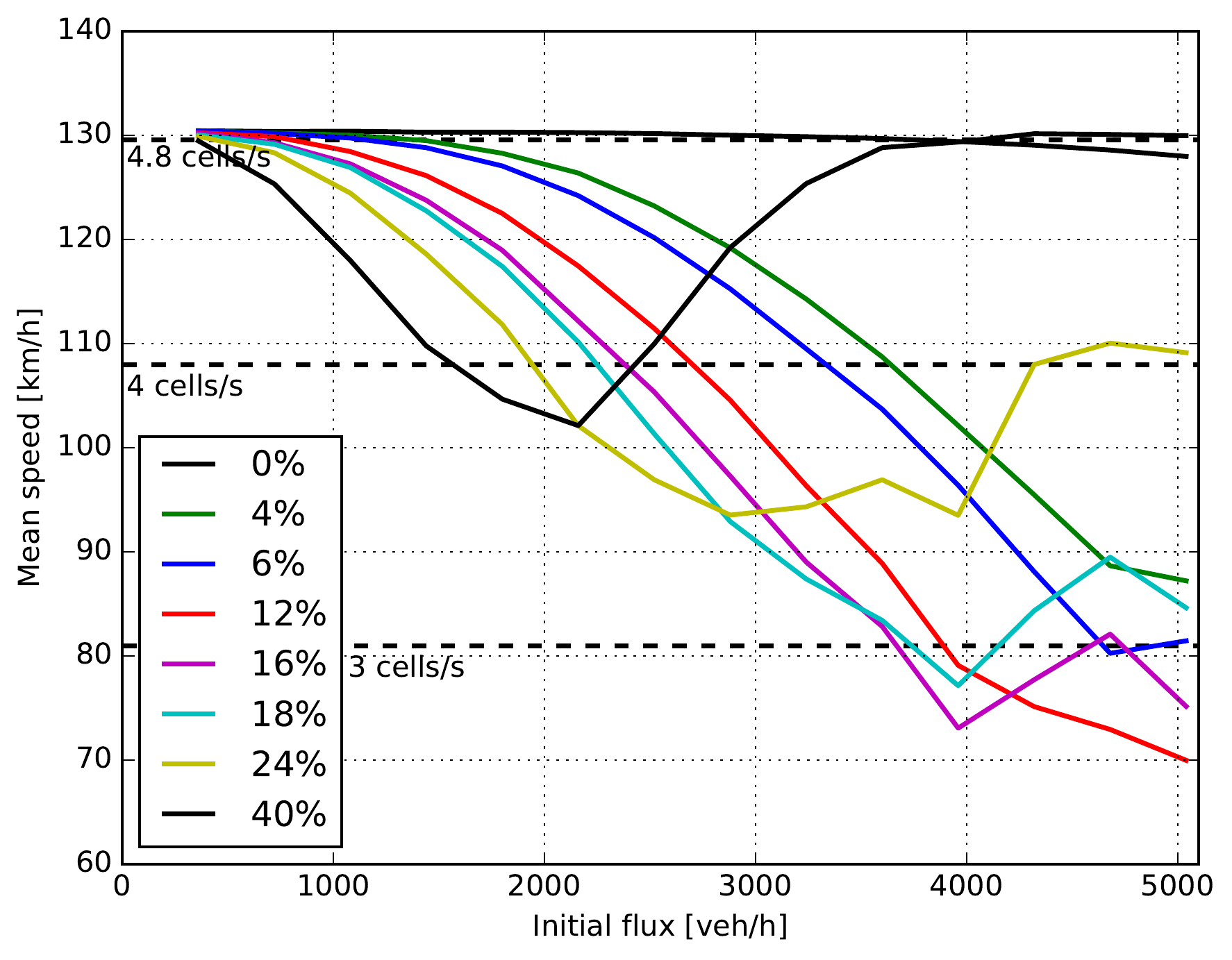}
\caption{}
\end{subfigure}
\hfill
\begin{subfigure}{0.49\textwidth}
\includegraphics[width=\textwidth]{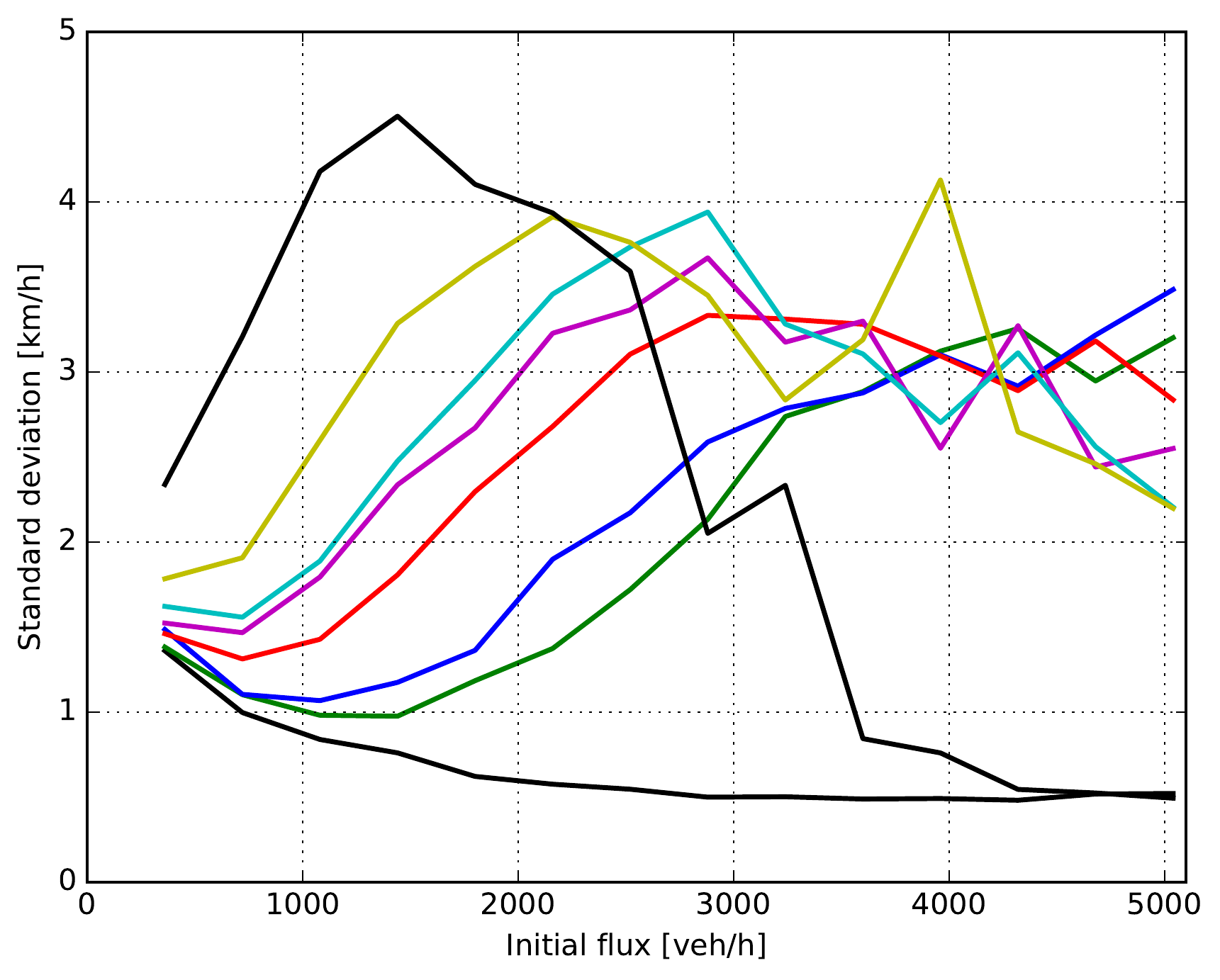}
\caption{\label{fig:D_Speed-S16}}
\end{subfigure}
\caption{\label{fig:unperturbed_speed}(a) Mean speed of fast vehicles and its (b) standard deviation at section 16 of S2, once the slope affects slow vehicles. Dashed lines in (a) represent the discrete values that the model allows as speeds in cells/s. We remember that 1 cell/s = 27 km/h. The top dashed line represents the maximum speed minus the rate of random noise affecting the speed at any given time step, i.e. 5 cells/s - 0.2 cells/s = 4.8 cells/s}
\end{figure}

The slope affecting slow vehicles in S2 starts in section 14. In Figure \ref{fig:unperturbed_speed} we observe the mean speed of fast vehicles at the left lane of section 16 (3.75km after the slope begins) with respect to the initial flux and the proportion of slow vehicles, with their corresponding standard deviation. As said in Section \ref{ssec:stdev}, for $r_1=0$ the standard deviation of speed diminishes as the initial flux increases, mainly because the mean speed does not suffer essential changes. In that sense, this is interpreted as the system ordering itself around a local mean speed. Nonetheless, this behavior is only observed for $r_1=0$. When slow vehicles appear in the system and form a mixed flux, the system transits to an unorganized state as the initial flux increases. The way this disorganization is done seems to have different and interesting behaviors which can be seen in Figure \ref{fig:D_Speed-S16}.

Only some of the proportions of slow vehicles are shown in Figure \ref{fig:unperturbed_speed}. This is to maintain the cleanness of the Figure and to focus in the relevant cases. Two different behaviors can be observed for $r_1>0$. For $r_1<0.16$ it is observed how the mean speed decreases as the initial flux increases, obtaining a local minimum between 4800 and 5000 veh/h of initial flux. Parallel to that, the standard deviation related increases. 
This is because the number of slow vehicles increases, blocking fast vehicles, which are forced to decrease their speed. As the number of slow vehicle increases, more fast vehicles are perturbed by them, resulting in the increase of the standard deviation.
When $0.16<leq r_1\leq 0.24$, mean speed bounces when the initial flux is equal 3960 veh/h after decreasing. Notice how, at this particular value of initial flux the mean speed meets a local minimum. After this value of initial flux, mean speed tends to the values that the model allows as speed. In particular, for $r_1=0.24$ we observe a transition phase to a synchronized flow. In this phase, the mean speed tends to a value which is not perturbed by the increase of the number of vehicles inside the system.

It is interesting to ask why, for certain proportion of slow vehicles, do mean speeds bounce after a certain initial flux. Given that initial flux determines the number of vehicles inside 
the highway, then for a certain total number of vehicles in a mixed flux, their transit is disorganized, leading to a decrease in the mean speed. However, given a certain proportion of slow vehicles ($>$16\%), fast vehicles are able to organize themselves to drive quicker. This is possible thanks to the set of rules imposed in Section \ref{ssec:cambios}.  

What happens between a free flux and a synchronization phase is also interesting to analyze. Taking the case of 24\% of slow vehicles and in a range between 0 veh/h and 2100 veh/h, the system is disorganizing itself as said before. On the other hand, when found in an initial flux of 3960 veh/h, a maximum standard deviation is found, meaning a maximum disorganization in the system. From there, the system abruptly organizes itself, described by an important decrease in the standard deviation around a higher mean speed. It is this maximum standard deviation and a consequent abrupt decrease what reflects a phase transition from a free flux to a synchronization phase.

The slope in S2 have important effects in the transit of the highway. The modifications described in Section \ref{ssec:modifications} do not change anything about the effect of gravity in heavy transportation. Thus, effective strategies must be found to optimize transit in the built infrastructure.

\subsection{\label{ssec:Perturbed}Perturbed system}

Ramps are considered, thus obtaining the complete system. Also, having this analysis done in two parts allow to know how much do ramps affect the system. Travel times are introduced. Results are presented for both senses S1 and S2 separately. 

\begin{figure}[ht]
\begin{subfigure}{0.49\textwidth}
\includegraphics[width=\textwidth]{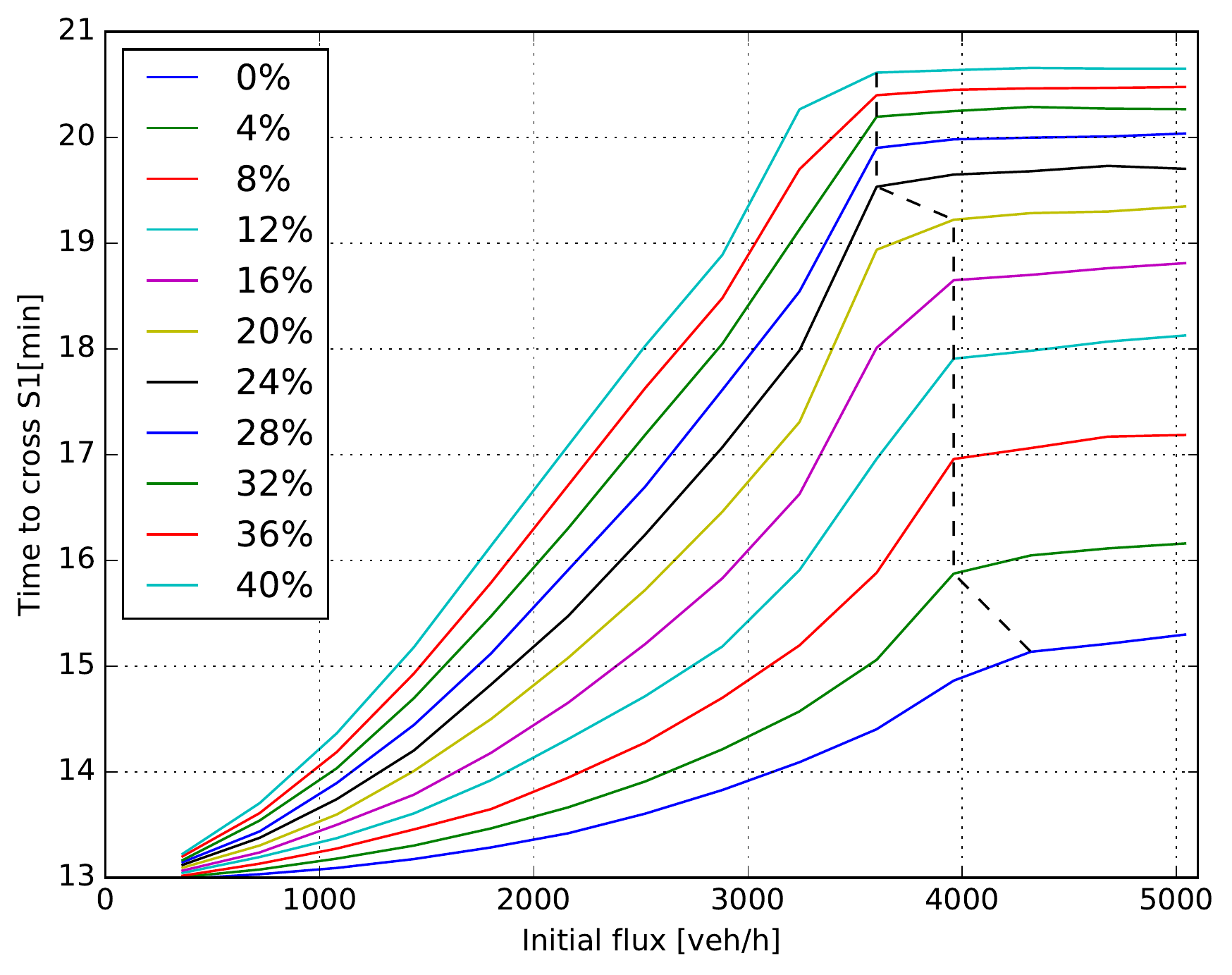}
\caption{\label{sfig:Times-S1-Original}}
\end{subfigure}
\hfill
\begin{subfigure}{0.49\textwidth}
\includegraphics[width=\textwidth]{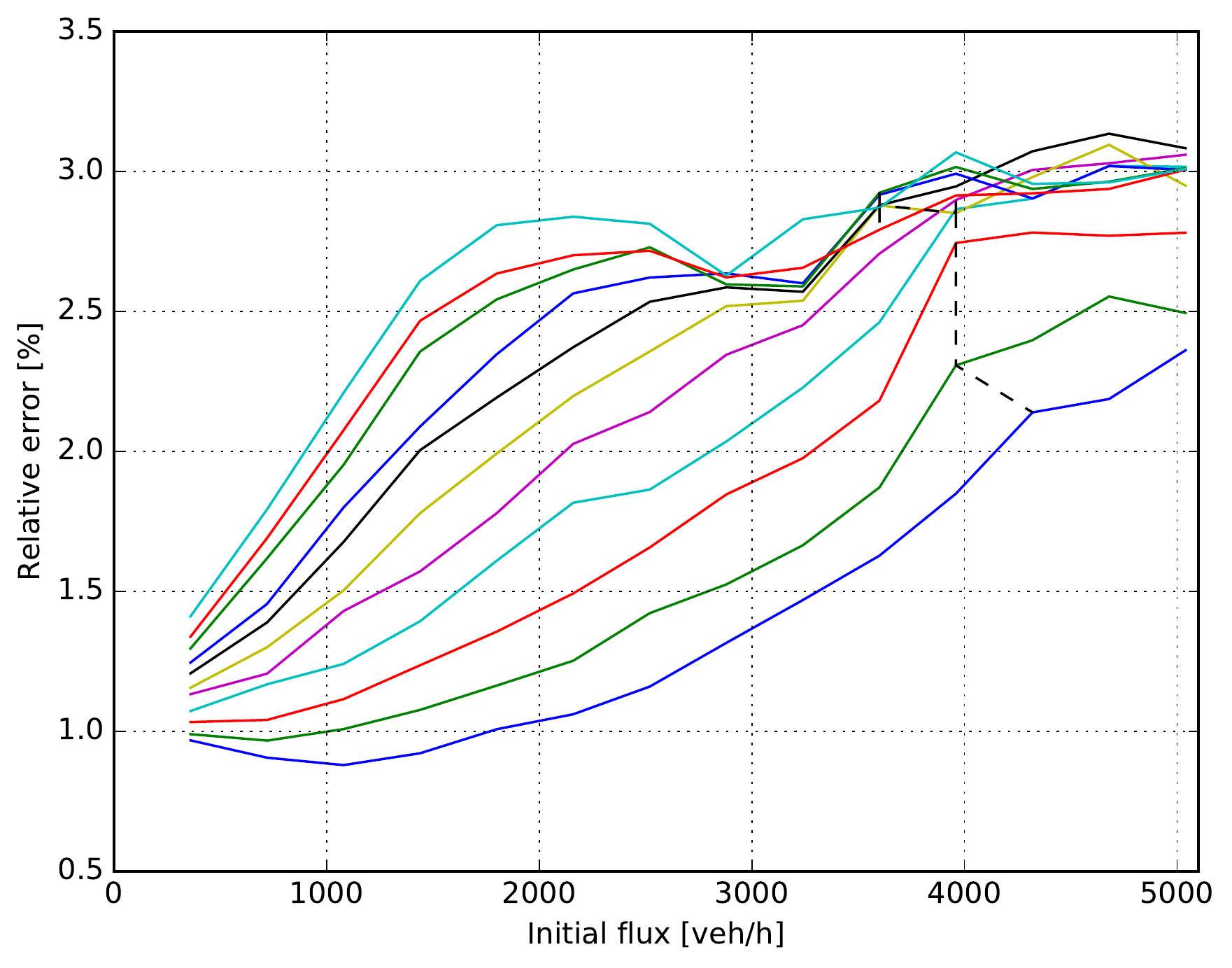}
\caption{\label{sfig:D_Times-S1-Original}}
\end{subfigure}
\caption{\label{fig:Times-S1-Original}(a) Travel times of fast vehicles and (b) its relative error for S1 Original. Each line represents a different proportion of slow vehicles, going from 0\% to 40\% in 4\% steps, starting with 0\% at the bottom and finishing at 40\% at the top. The dashed line represents the phase transition. }
\end{figure}

\subsubsection{Results for S1}

In Figure \ref{sfig:Times-S1-Original}, travel times for fast vehicles in S1 before the governmental works (S1 Original) are shown with respect to the proportion of slow vehicles and the initial flux. Each bold line represents a fixed proportion of slow vehicles, going from 0\% to 40\% in steps of 4\%. In the case of travel times, we do not have standard deviations as travel times are indirect measurements computed using the mean speed, following Eq. \ref{eq:Times}. Instead of standard deviation, we have errors that related to the mean speed and its standard deviation. In Figure \ref{sfig:D_Times-S1-Original}, the relative errors are plotted. 

Relative error, in comparison to standard deviation, does not give information about transition phases. This can be seen in Figure \ref{fig:Times-S1-Original}. A dashed line is drawn in order to delimit two different phases in the system.  
The behavior to the left of the dashed line refers to a gaseous or free flux regime, where the average speed decreases as the initial flux increases as seen in Section \ref{ssec:Unperturbed}, resulting in an increase of the travel time. The behavior to the right, on the other hand, refers to a synchronization. This results in an unperturbed travel time by the initial flux. This synchronization comes also with a maximum standard deviation which is negligibly perturbed by the initial flux.
However, comparing to the results in Section \ref{ssec:Perturbed}, the relative error does not reflect the transition phase with a peak of transition phase, but rather has a smooth transition phase. This contrasts to the fact that the studied model does not present soft phase transitions\cite{larragadelrio2005}. 

\begin{figure}[ht]
    \centering
    \includegraphics[width=0.5\textwidth]{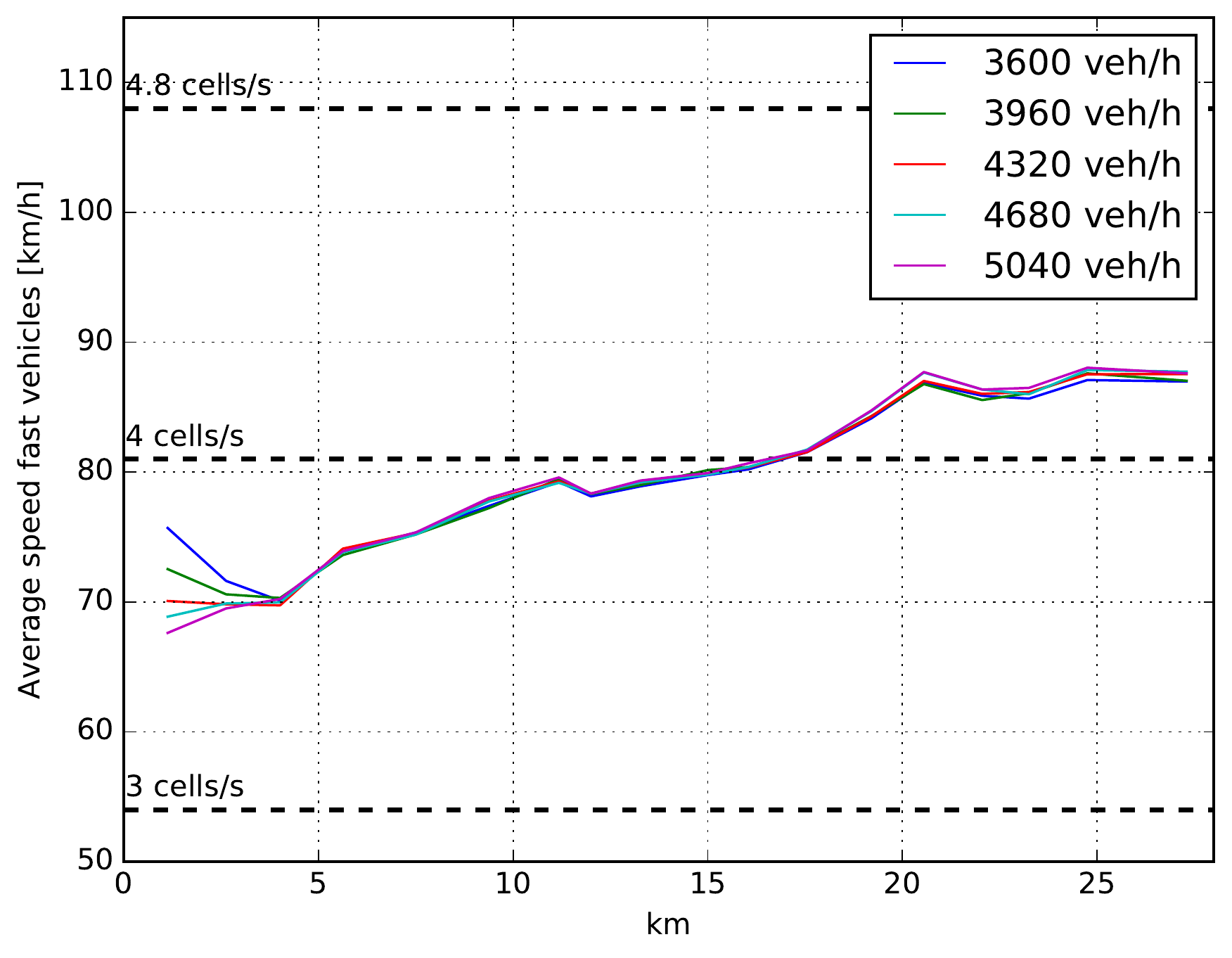}
    \caption{\label{fig:Synchro}Average speed of fast vehicle along S1 Original, for initial fluxes between 3600 veh/h and 5040 veh/h.}
\end{figure}

In order to understand what happens in this synchronization phase, the average speed along the highway is plotted in Figure \ref{fig:Synchro} for 40\% of slow vehicles and an initial flux between 3600 veh/h and 5040 veh/h where the travel time is 20.6 min. As it can be seen, average speeds only differ in the first measure at the beginning of the highway. This comes from the fact that the flux is supposed to initiate at the beginning of the highway, while in the physical highway flux initiate a dozen of kilometers before. The minimal change in average speed results in a constant travel time. \\

\begin{figure}[ht]
\begin{subfigure}{0.49\textwidth}
\includegraphics[width=\textwidth]{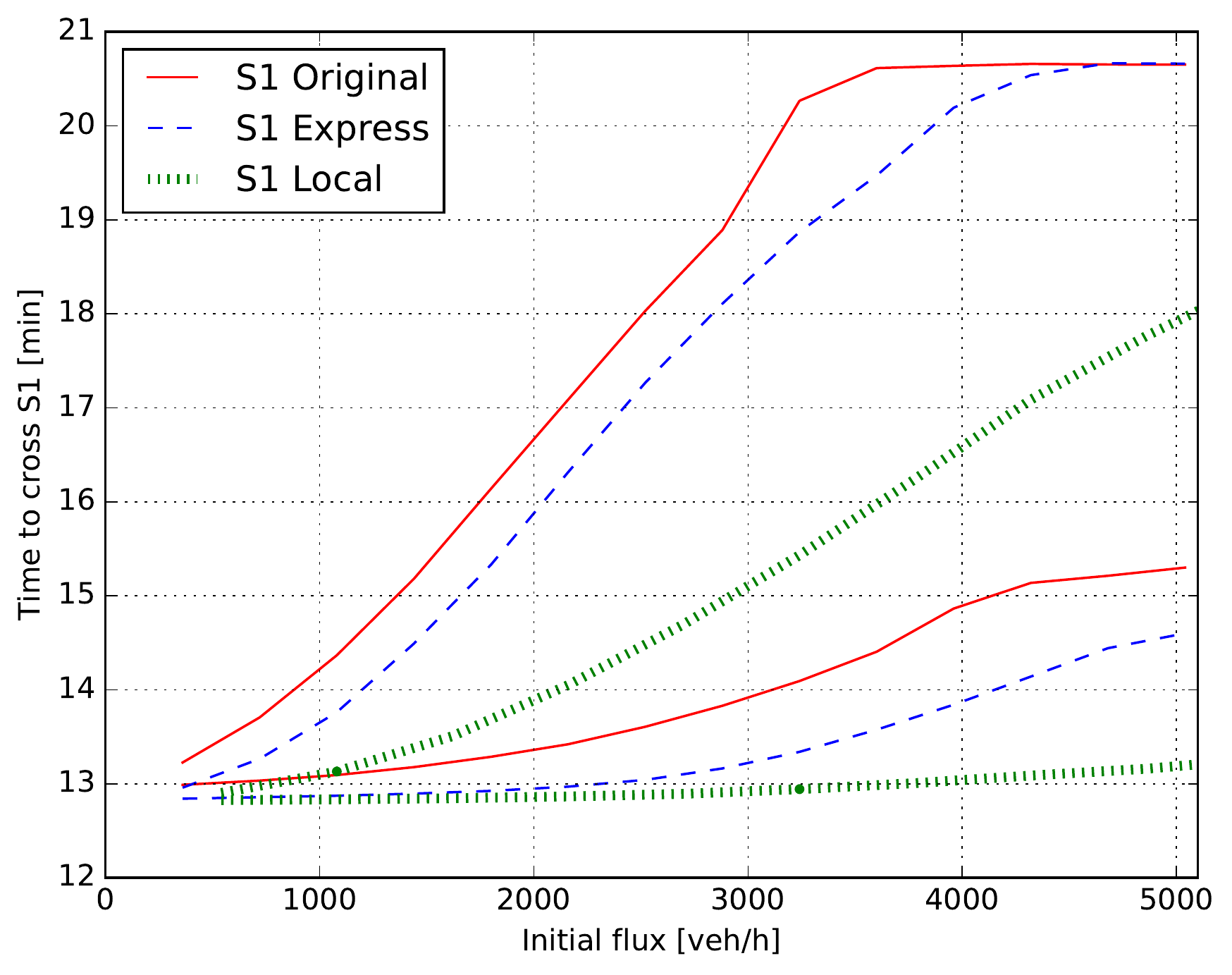}
\caption{\label{sfig:S1-Times}}
\end{subfigure}
\hfill
\begin{subfigure}{0.49\textwidth}
\includegraphics[width=\textwidth]{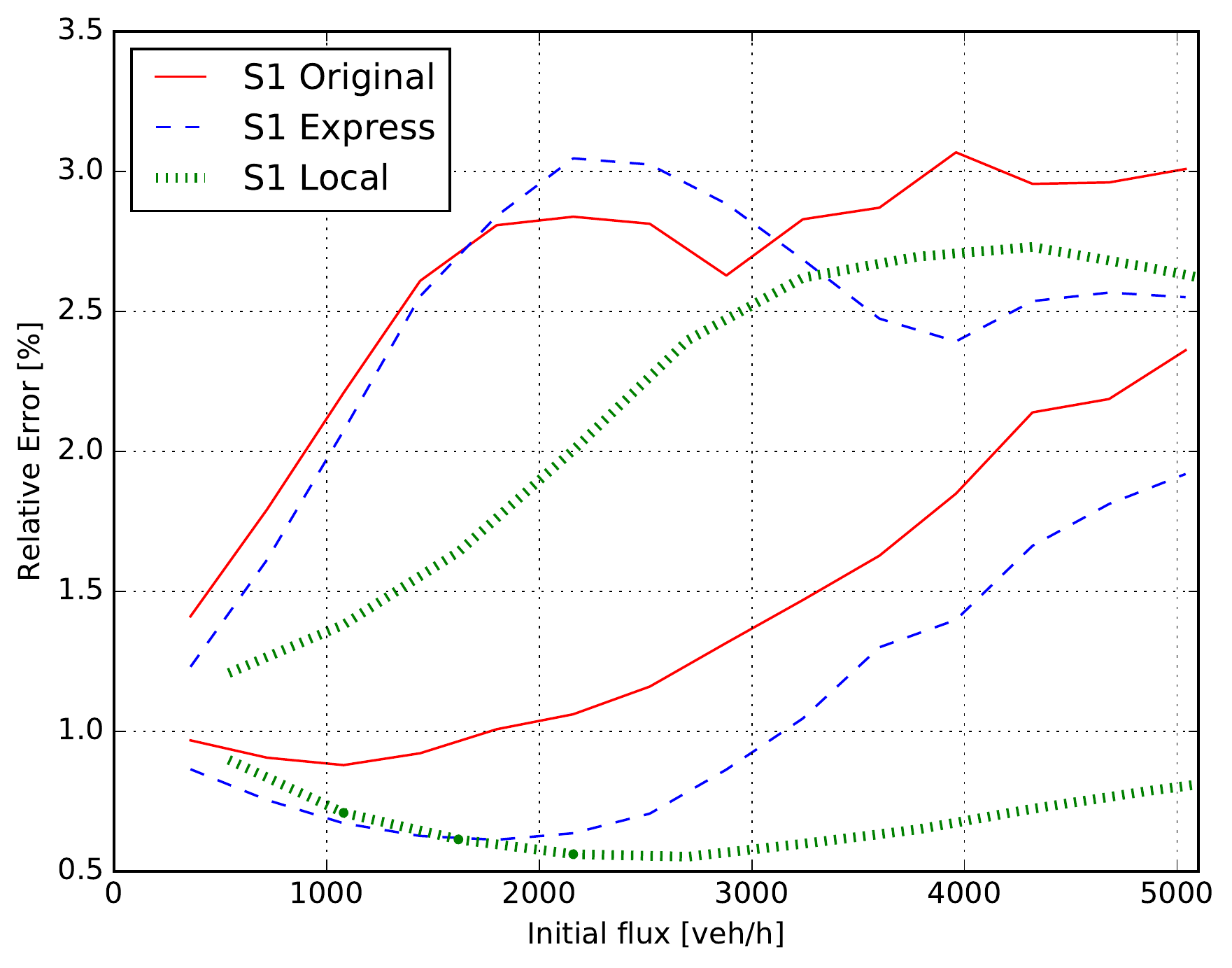}
\caption{\label{sfig:D_S1-Times}}
\end{subfigure}
\caption{\label{fig:S1-Times}(a) Travel times of fast vehicles and (b) its relative error for S1. Only two lines per system are presented, corresponding to 0\% and to 40\% of slow vehicles. These lines serve as envelops to the general behavior.}
\end{figure}

The qualitative behavior observed for S1 Original in Figure \ref{fig:Times-S1-Original} is repeated for S1 Express and for S1 Local. This allows to directly compare travel times of the three systems and to have a more complete picture of what the different governmental works produce. Results are shown in Figure \ref{fig:S1-Times}. Instead of plotting every different percentage of slow vehicles, only the maximums are present (0\% and 40\%), representing the envelops.

From Figure \ref{sfig:S1-Times} it is observed how the best travel times are obtained at S1 Local, having an increasing behavior as the initial flux is increased and no synchronization phase. On the other hand, S1 Express has better times than S1 Original but keeps the same transition to a synchronized phase. This transition is done more smoothly than in S1 Original, reaching the synchronized travel time after an initial flux of 4500 veh/h. 

Looking at Figure \ref{sfig:D_S1-Times}, it is noticeable how S1 Local presents the best travel times. 
The main difference between the latter system and the others is the addition of a third lane during the 14.5 first kilometers. The existence of this third lane allows freeing the extreme left one of slow vehicles, thus having space with higher average speed, decreasing travel times. However, whereas the extreme left lane exists during the 27.3 kilometers of S1 Local, the extreme right only exists for the 14.5 first kilometers. The result of this is that all vehicles being in the extreme right lane must merge into the second if they want to keep in S1, explaining the broadening of the relative error in Figure \ref{sfig:D_S1-Times}.

Also, the broadening of the relative error of S1 Express is explained by the fact that, as Figure \ref{fig:rampsS1-2} shows, the deletion of 8 ramps provokes a net flow of 500 veh/h at the end of the Express Pass. For small initial fluxes, the ramp perturbs the flow inside the highway. As the initial flux increases, the number of vehicles is such that the flux in the ramps has a weaker effect in the flux inside the highway.\\

From the present results, an easy-to-implement strategy to optimize the traffic can be presented. If the traffic of slow vehicles is prohibited in S1 Express, then travel times would be reduced significantly. The effects of this strategy can be seen in Figure \ref{fig:S1-Times}. In S1 Original the interval of travel times goes from 13 min $\pm$ 1\%  to 20.7 min $\pm$ 3\%, which is the same interval of times in the modified highway (S1 Local + S1 Express). However, if slow vehicles cannot transit in S1 Express, then only the lower part of the envelop of S1 Express must be taken in Figure \ref{fig:S1-Times}. The interval of travel times will then be from 13 min $\pm$ 1\% to 18 min $\pm$ 2.6\%, meaning a reduction of almost 10\% in travel times. 

\subsubsection{Results for S2}

\begin{figure}[ht]
\begin{subfigure}{0.49\textwidth}
\includegraphics[width=\textwidth]{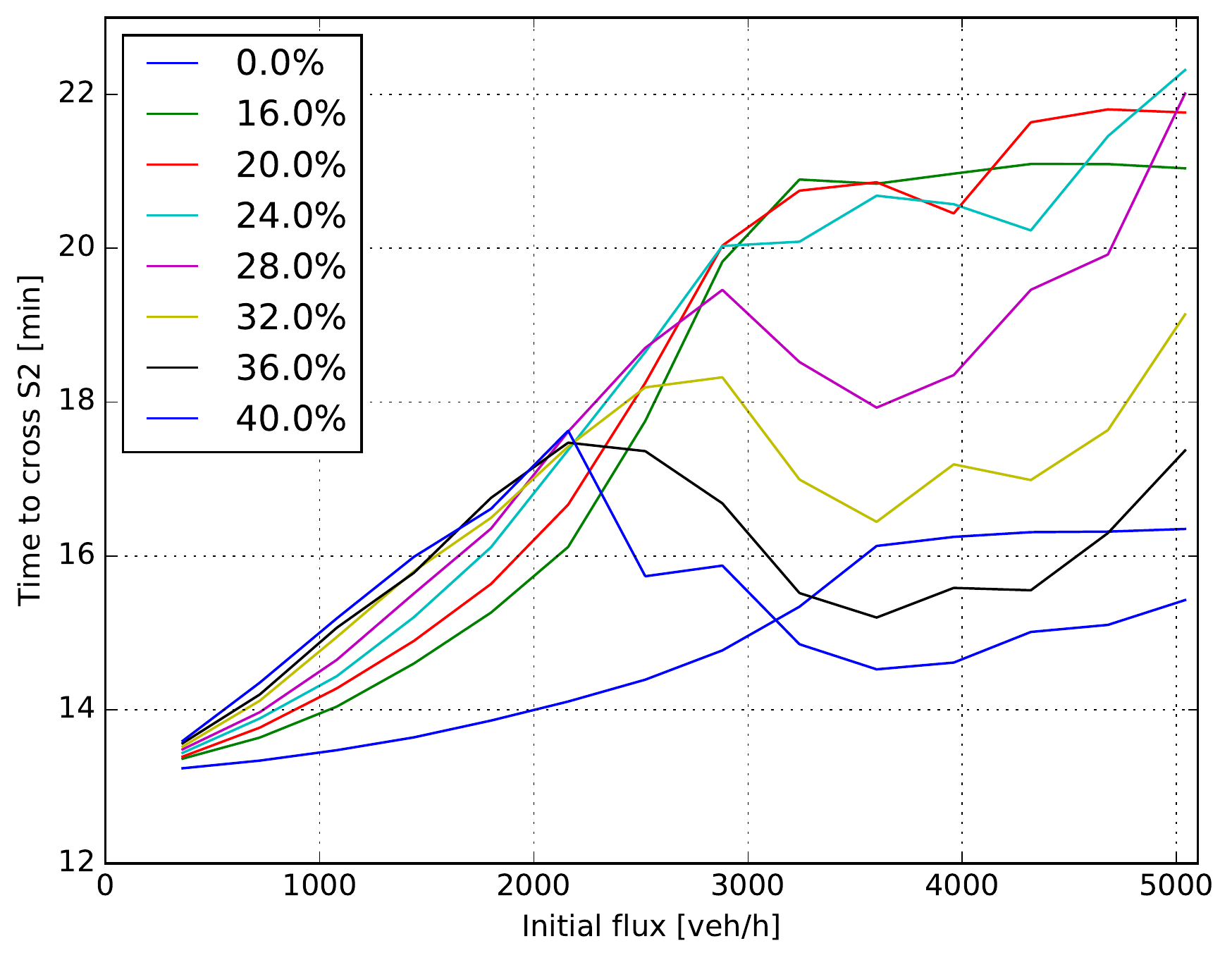}
\caption{\label{sfig:Times-S2-Original}}
\end{subfigure}
\hfill
\begin{subfigure}{0.49\textwidth}
\includegraphics[width=\textwidth]{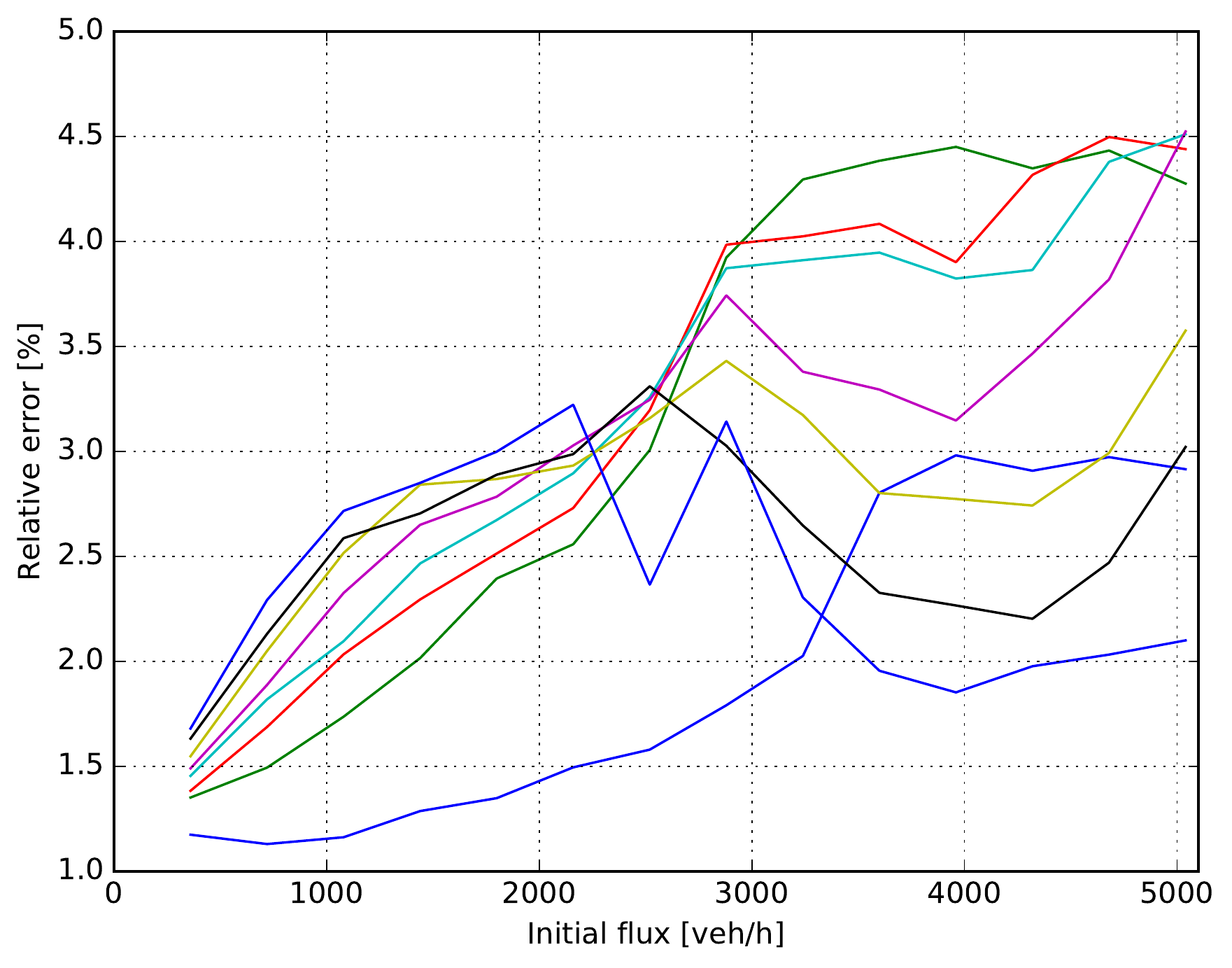}
\caption{\label{sfig:D_Times-S2-Original}}
\end{subfigure}
\caption{\label{fig:Times-S2-Original}(a) Travel times of fast vehicles and (b) its relative error for S2 Original.}
\end{figure}

S2 presents a different behavior with respect to S1. Results are shown in Figure \ref{fig:Times-S2-Original}. In this case, the behavior of travel times of fast vehicles are not as simple as in S1, and only some of them are presented to maintain cleanness. Between 0\% and 16\% of slow vehicles, travel times behave similar to S1, increasing as the initial flux increases, to then become constant, entering into a synchronization phase. However, once this proportion is passed, the synchronization phase is broken, observing a second kind of transition phase. This time, the transition is not only achieved varying the initial flux, but also increasing the proportion of slow vehicles. 
It is important to notice what happens between 16\% and 24\% of slow vehicles, as it is in this interval where the transition phase happens. From a stable travel time at 16\%, a destabilization starts to happen at 20\% and 24\%, finishing with a complete different behavior of travel times at 28\% of slow vehicles. 
It is in this transition phase where the maximum travel times and relative errors are found. It is undesirable to be driving during this transition phase, as it involves  a higher probability of having an accident.

\begin{figure}[ht]
\begin{subfigure}{0.49\textwidth}
\includegraphics[width=\textwidth]{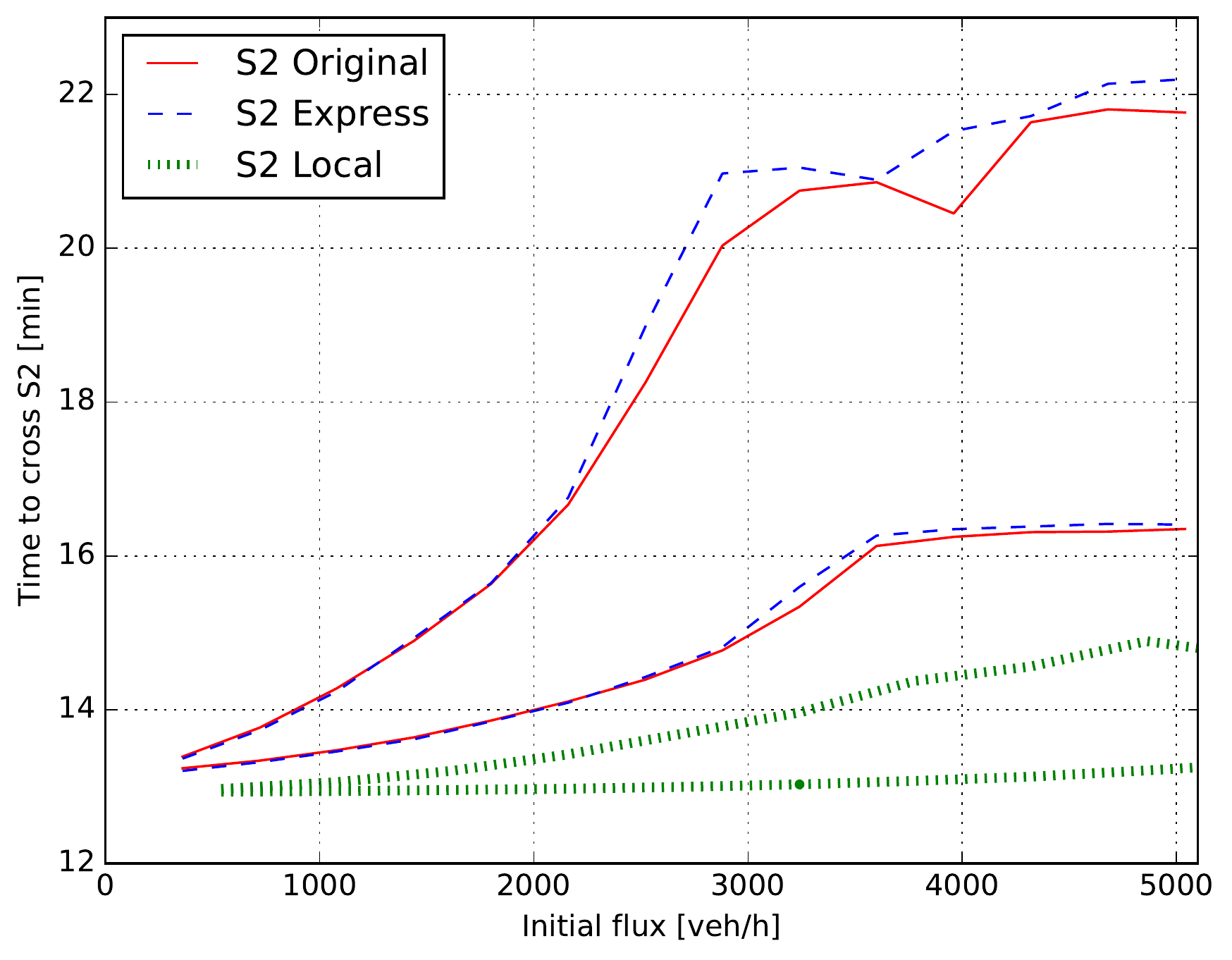}
\caption{\label{sfig:S2-Times}}
\end{subfigure}
\hfill
\begin{subfigure}{0.49\textwidth}
\includegraphics[width=\textwidth]{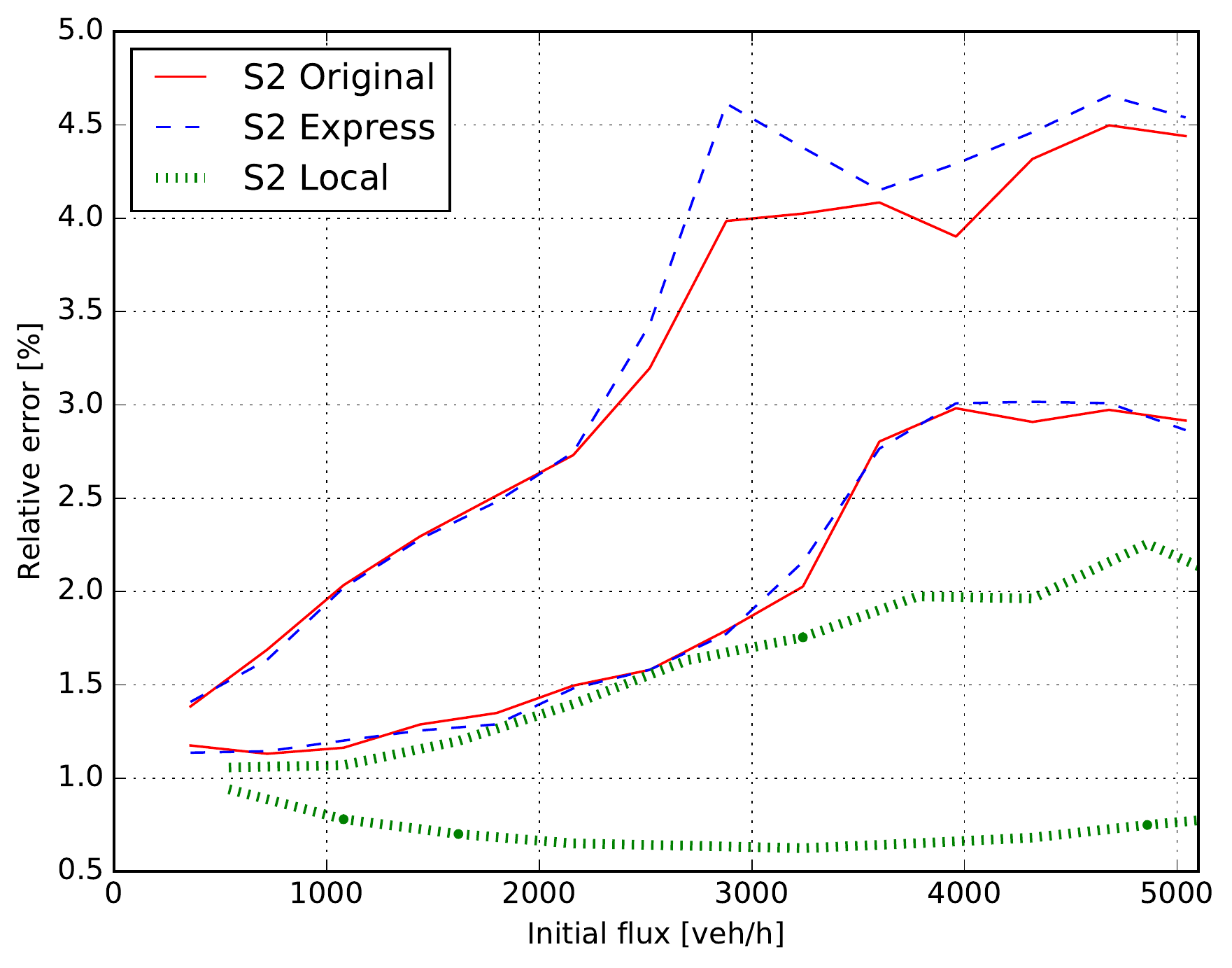}
\caption{\label{sfig:D_S2-Times}}
\end{subfigure}
\caption{\label{fig:S2-Times}(a) Travel times of fast vehicles and (b) its relative error for S2. Only two lines per system are presented, corresponding to 0\% and to 20\% of slow vehicles. These lines serve as envelops to the general behavior.}
\end{figure}

Similar to S1, Figure \ref{fig:S2-Times} presents the envelops of the travel times and its standard deviation for the three systems of S2. However, in this case, the envelops are shown by the lines corresponding to 0\% and 20\% of slow vehicles. 
S2 Express presents the same general behavior of S2 Original which has been already discussed. S2 Local, on the other hand, does not present transitions of any kind, leaving the system in a gaseous state. It is here that the smaller travel times and standard deviations are found. Indeed S2 Local seems to follow a complete different dynamics than S2 Original and S2 Express. The main reason of this is the presence of the third lane during the second half of S2, where the slopes affects the slow vehicles. As in S1, the third lane allows fast vehicles to have a space to accelerate and go to a higher speed. Furthermore, for S1 Local it is observed a broadening of travel times and its relative error in Figure \ref{fig:S1-Times}, while for S2 Local, travel times and the relative error remains do not see such and increase. 
The main factor for this to happen is the location of the third lane with comparison of the global transit in the highway. While in S1 Local, the third lane finishes at the 14.5 km of the modified part, and thus creating a bottleneck in both S1 Express and S1 Local, for S2 Local the third lane has the opposite effect. If there is a very dense and slow transit in the first sections of S2, then this transit will disperse when the third lane is added.
In this sense, the third lane in S2 Local buffers the topographical effect of the slope in slow vehicles

In this case the same strategy as in S1 can be implemented, obtaining even better results. The introduction of the Local system  provoked a different dynamics in the highway, reducing the travel times to 15 min $\pm$ 2.25\% in the most critical case of 5000 veh/h and 40$\%$ of slow vehicles, meaning a reduction of 32$\%$ in travel times. By itself, S2 Local made an important improvement. However, if no strategy is implemented, the driver does not experiment any change in terms of time when crossing S2 Express than when he crossed S2 Original. If all slow vehicles are forced to be in S2 Local (which does not have an important effect as just said), travel times in S2 Express are reduced to an interval between 13 min $\pm$ 1.1\% and 16.3 min $\pm$ 3.0\%, meaning a reduction of 26\% in travel times. 

\section{\label{sec:Conclusions}Conclusions}

The Mexican Government realized several works to the Cuernavaca bypass in order to improve the mobility of the inhabitants of the metropolitan zone. The availability of traffic data gives the opportunity to analyze not only the bypass before the works but to project the behavior of the highway afterwards. \\

To analyze and detect phase transitions, a particular emphasis is put on the standard deviation of the mean average speed. A presentation was done to use the standard deviation as a measure of order, useful when other variables such as entropy are not available. The radical fluctuations of standard deviation allowed to detect phase transitions from free flux regime (or gaseous state) to congested regimes (or liquid state) and to a synchronized phase where the macroscopic measure does not change as the number of vehicles in the system increases. \\

However, as the government works results in three different systems (one past, two present), the analysis of every one becomes unpractical. In that sense, travel times are computed. This variable allows to have a macroscopic measure analyzing the system in a whole, whilst being attainable and intuitive for physicist, transit analysts and engineers. This analysis allows to detect not only gaseous and liquid states, but also synchronous phase in the different systems. \\

The analysis done in Section \ref{sec:Results} shows how, without any specific strategy, the maximum travel times in the modified highway stay the same, while the minimum travel times improve. In order to actually make a profit of the modifications done by Government, an easy-to-implement strategy must be done. Restricting the transit in the Express Pass to slow vehicles. It is only in that sense that the modifications take sense. 
The analysis also shows how topographical elements present in the bypass affect the transit in the highway, more specifically in the South-North sense. It is here where an unstable free flux regime is obtained, finding the highest values for travel times and standard deviations. This unstable regime, which is found for a proportion of slow vehicles between 16\% and 24\% is not only the most inconvenient in terms of travel times, but also the most dangerous, as the high standard deviations reflects a highest probability of having an accident. However, the addition of a third lane in the ``Local'' system allows to buffer the topographic effect, resulting in an important decrease of travel times of up to 32\%.\\

Also, and more importantly, the study presented here shows how physical interpretations of many body systems of a non-physical nature can be successfully applied to describe observables which depend of human behavior. But most importantly, to study and interpret transition phases in human systems using standard deviation as a measure of order.\\

Official data release by the government is of enormous usefulness in order to analyze and create better strategies for the common wellness. In that sense, transparency policy must move towards a regime where data is measured with great care and rigor and released to the general public (always respecting privacy) in order to obtain critical analysis from the scientific and industrial society.
\section*{Acknowledgments}
The authors would like to thank the Centre for Complexity Sciences (C3-UNAM) for the use of their computational cluster ``Borromeo'' to execute the simulations for the present work.

\end{document}